\documentclass[fleqn,usenatbib]{mnras}

\usepackage[T1]{fontenc}

\DeclareRobustCommand{\VAN}[3]{#2}
\let\VANthebibliography\thebibliography
\def\thebibliography{\DeclareRobustCommand{\VAN}[3]{##3}\VANthebibliography}

\usepackage{graphicx}	
\usepackage{amsmath}	
\usepackage{amssymb}	

\usepackage{ulem} 

\usepackage{newtxtext,newtxmath}

\newcommand{\Msun}{\,{\rm M}_{\sun}}

\newcommand{\Gyr}{\,{\rm Gyr}}
\newcommand{\kpc}{\,{\rm kpc}}

\newcommand{\Mc}{M_{\rm c,*}}

\title[Upper truncation mass of the GC mass function]{The physics governing the upper truncation mass of the globular cluster mass function}

\author[M. E. Hughes]{
Meghan E. Hughes$^{1}$\thanks{E-mail: m.hughes1@2013.ljmu.ac.uk},
Joel L. Pfeffer$^{2,1}$,
Nate Bastian$^{1,3,4}$,
Marie Martig$^{1}$,
\newauthor
J. M. Diederik Kruijssen$^{5}$,
Robert A. Crain$^{1}$,
Marta Reina-Campos$^{6,7}$,
\newauthor
Sebastian Trujillo-Gomez$^{5}$
\\
$^{1}$Astrophysics Research Institute, Liverpool John Moores University, 146 Brownlow Hill, Liverpool L3 5RF, UK\\
$^{2}$International Centre for Radio Astronomy Research (ICRAR), M468, University of Western Australia, 35 Stirling Hwy, Crawley, WA 6009, Australia \\
$^{3}$Donostia International Physics Centre (DIPC), Paseo Manuel de Lardizabal 4, 20018 Donostia-San Sebastian, Spain \\
$^{4}$IKERBASQUE, Basque Foundation for Science, E-48013 Bilbao, Spain\\
$^{5}$Astronomisches Rechen-Institut, Zentrum f\"{u}r Astronomie der Universit\"{a}t Heidelberg, M\"{o}nchhofstra{\ss}e 12-14, 69120 Heidelberg, Germany\\
$^{6}$Department of Physics \& Astronomy, McMaster University, 1280 Main Street West, Hamilton, L8S 4M1, Canada\\
$^{7}$Canadian Institute for Theoretical Astrophysics (CITA), University of Toronto, 60 St George St, Toronto, M5S 3H8, Canada\\
}

\date{Accepted XXX. Received YYY; in original form ZZZ}

\pubyear{2015}

\begin{document}
\label{firstpage}
\pagerange{\pageref{firstpage}--\pageref{lastpage}}
\maketitle

\begin{abstract}
The mass function of globular cluster (GC) populations is a fundamental observable that encodes the physical conditions under which these massive stellar clusters formed and evolved. The high-mass end of star cluster mass functions are commonly described using a Schechter function, with an exponential truncation mass $\Mc$. For the GC mass functions in the Virgo galaxy cluster, this truncation mass increases with galaxy mass ($M_{*}$). In this paper we fit Schechter mass functions to the GCs in the most massive galaxy group ($M_{\mathrm{200}} = 5.14 \times 10^{13} \Msun$) in the E-MOSAICS simulations. The fiducial cluster formation model in E-MOSAICS reproduces the observed trend of $\Mc$ with $M_{*}$ for the Virgo cluster. We therefore examine the origin of the relation by fitting $\Mc$ as a function of galaxy mass, with and without accounting for mass loss by two-body relaxation, tidal shocks and/or dynamical friction. 
In the absence of these mass-loss mechanisms, the $\Mc$-$M_{*}$ relation is flat above $M_* > 10^{10}\Msun$. It is therefore the disruption of high-mass GCs in galaxies with $M_{*}\sim 10^{10} \Msun$ that lowers the $\Mc$ in these galaxies. High-mass GCs are able to survive in more massive galaxies, since there are more mergers to facilitate their redistribution to less-dense environments. The $\Mc-M_*$ relation is therefore a consequence of both the formation conditions of massive star clusters and their environmentally-dependent disruption mechanisms.
\end{abstract}




\section{Introduction}

The luminosity, or mass, function of globular clusters (GCs) is one of the most fundamental observables to link the formation of star clusters to `normal' star formation processes. 
Unlike the near power-law mass functions of young star clusters \citep[or young `massive' clusters, YMCs; e.g.][]{Whitmore1995, Miller1997, Larsen2002}, generally considered to be the young equivalents of today's old GCs \citep[see reviews by][]{PortegiesZwart2010, Kruijssen2014, Forbes2018, Krumholz2019}, the luminosity/mass functions of GCs appear peaked at similar magnitudes in all environments, corresponding to a turnover mass of $M_{\mathrm{TO}} \approx 2 \times 10^5 \Msun$ \citep[e.g.][]{Harris2001, Brodie2006, Jordan2007}.

If we take the view that the GCs we observe today formed in a similar way to the YMCs forming in the local Universe then the mass function must have been transformed in some way at the low mass end. To explain this we have to take into account that the old GCs we observe at $z=0$ are the surviving population of those initially formed. Therefore a strong possibility to explain the transformation in the mass function is a preferential destruction of low mass star clusters by dynamical processes (e.g. \citealt{Okazaki1995, Baumgardt1998, Vesperini1998, Vesperini2003, Fall2001, deGrijs2003, Goudfrooij2004, Gieles2006b, Elmegreen2010b, Kruijssen2012b}).

Traditionally, the GC luminosity/mass function has been modelled as a Gaussian or log-normal distribution to fit the peak (or turnover) of the luminosity function \citep[e.g.][]{Hanes1977, Harris1979, van_den_Bergh1985, Harris2001}. 
The turnover mass (or luminosity) varies only weakly with galaxy mass \citep{Jordan2007, Villegas2010}.
The high mass end of the GC mass function (i.e. above the turnover mass) has also been shown to be well fit by a power-law function with an index around $-1.7$ to $-2$ \citep{Surdin1979, Racine1980, Harris1994, Durrell1996, Kissler-Patig1996}, surprisingly close to that observed for YMCs \citep[$\approx -2$ with some galaxy-to-galaxy variations,][]{Whitmore1995, Miller1997, Larsen2002, Zhang1999, Bik2003, McCrady2007, Chandar2010, Whitmore2014, Messa2018a}.

However, at the very high mass end there is evidence for a further steepening of the cluster mass function, for both GCs \citep{Harris1994, Burkert2000} and YMCs \citep{Whitmore1999, Larsen2002, Larsen2009, Gieles2006, Bastian2012, Adamo2015M83, Adamo2017, Adamo2020, Johnson2017, Messa2018a}.
This steepening, or `truncation', is generally fit by a \citet{Schechter1976} function, i.e. a power law with an exponential truncation ($\Mc$).
For GC mass functions this may be modified in the form of an `evolved' Schechter function in order to fit both the high-mass truncation and turnover at lower masses \citep{Jordan2007}.
In the case of YMCs, the truncation may vary with local environment (namely the star formation rate surface density, \citealt{PortegiesZwart2010, Adamo2015M83, Adamo2020, Johnson2017, Messa2018b}; though see \citealt{Mok2019} for evidence for a constant truncation mass).
\citet{Reina-Campos2017} predict that the high mass end of the GC mass function does not depend on the absolute star formation rate, but instead is set by a combination of galactic dynamics and stellar feedback, resulting in an effective scaling with the gas and star formation rate surface densities. It is only when accounting for the interplay between both mechanisms that they can reproduce the observed trends of $\Mc$ with galactocentric radius.

In the case of GCs, fewer studies have investigated systematic trends of the upper truncation with galaxy properties.
\citet{Jordan2007} analysed the luminosity and mass function of evolved GCs observed by the ACS Virgo Cluster Survey to investigate the dependence of the GC luminosity and mass function on galaxy stellar mass. They find that the luminosity function of the GCs shows a decreasing $\Mc$ value with decreasing galaxy mass. They argue that the behaviour at the high mass end of the GC mass function is a consequence of systematic variations of the initial cluster mass function rather than long-term dynamical evolution.

In this work, we compare the relation between $\Mc$ and galaxy mass in the E-MOSAICS simulations (\citealt{Pfeffer2018,Kruijssen2019a}; Crain et al. in prep.) to that found by \citet{Jordan2007}. The E-MOSAICS simulations trace the formation and evolution of GC populations alongside galaxy formation and evolution, and therefore enable us to investigate the impact of various galaxy properties on the resulting GC observables. Specifically for this work, we can explore how GC formation environment and GC mass loss play a role in initialising and evolving the GC mass function. The E-MOSAICS simulations have been shown to reproduce and provide an explanation for a range of observed properties of both young and old GCs, such as the existence of a `blue tilt' in GC populations \citep{Usher2018}, as well as the fraction of disrupted GC stars in the bulge \citep{Hughes2020} and the halo \citep{ReinaCampos2018, ReinaCampos2020} of the Milky Way. The simulations have shown that the diversity in age-metallicity relations of Milky Way-mass galaxies results from different assembly histories and can therefore be used to infer such assembly histories \citep{Kruijssen2019a,Kruijssen2019b,Kruijssen2020}. The simulations have also been shown to reproduce the observed kinematics of the Galactic GC population \citep{TrujilloGomez2021} and have been used to conclude that GCs associated with stellar streams will be, on average, younger than the GC population not associated with a stellar stream \citep{Hughes2019}, a result subsequently confirmed through observations of stellar streams in the halo of M31 \citep{Mackey2019}. Finally, \citet{Pfeffer2019YC}  showed that the simulations reproduce the properties of young cluster populations, and the simulations were subsequently used to predict when and where GCs formed \citep{ReinaCampos2019,Keller2020}. By comparing the simulation outputs to the observations of \citet{Jordan2007}, this work will serve as another test that YMCs and ancient GCs share the same formation mechanism.
However, previous work showed the E-MOSAICS simulations do not reproduce the observed GC turnover mass, most likely due to under-disruption of low-mass GCs in the simulations \citep[see][]{Pfeffer2018, Kruijssen2019a}. Therefore in this work we focus on the high-mass end of the GC mass function (i.e. the exponential truncation $\Mc$).

This paper is organised as follows. In Section \ref{2}, we outline the aspects of the E-MOSAICS simulations important for this work. In Section \ref{3}, we describe the observational data and compare them to the simulations. In this section, we also justify our choice to fit \citet{Schechter1976} functions instead of a single power-law function. Section \ref{4} investigates the impact of considering alternative cluster formation scenarios on the mass functions. Section \ref{5} describes how the mass function changes when including the initial masses of GCs that have been completely disrupted. Finally, in Section \ref{6} we present our conclusions.

\section{Simulations} \label{2}
 
The E-MOSAICS suite of simulations \citep{Pfeffer2018,Kruijssen2019a} couples the MOSAICS \citep{Kruijssen2011,Pfeffer2018} model for star cluster formation and evolution to the EAGLE \citep{Schaye2015,Crain2015} model for galaxy formation. This enables the simulations to follow simultaneously the formation and evolution of star clusters and their parent galaxies in a cosmological context. This work uses the E-MOSAICS $34.4^3$ comoving Mpc$^3$ periodic volume (Crain et al. in prep.), first featured in \citet{Bastian2020}. The E-MOSAICS model is the same as described for the E-MOSAICS zoom-in simulations \citep{Pfeffer2018,Kruijssen2019a}. The simulations have gas particles with initial masses $m_{\mathrm{g}} = 2.25 \times 10^5 \Msun$.\par

EAGLE is a set of  hydrodynamical simulations of the formation of a cosmologically representative sample of galaxies in a $\Lambda$CDM cosmogony, meaning that a wide range of galaxy environments are sampled. The simulations include sub-grid radiative cooling \citep{Wiersma2009}, star formation \citep{Schaye2008}, stellar feedback \citep{DallaVecchia2012}, chemical evolution \citep{Wiersma2009}, gas accretion onto, and mergers of,  super massive black holes (BHs) \citep{Rosas-Guevara2015} and active galactic nuclei (AGN) feedback \citep{Booth2009}. The standard resolution EAGLE simulations yield a galaxy stellar mass function that reproduces the observed function to within 0.2 dex over the well-sampled and well-resolved mass range. 
For a full description of the models, see \cite{Schaye2015}.\par

The EAGLE model has been shown to reproduce many important galaxy properties, such as galaxy masses, sizes, luminosities and colours \citep{Furlong2015,Furlong2017,Trayford2015}, as well as their cold gas properties (e.g. \citealt{Lagos2015,Crain2017}) and cosmic star formation rate density \citep{Furlong2015}. The wide range of reproduced galaxy properties across a wide range of galaxy masses makes the EAGLE model ideal for comparing with observed galaxy properties.\par

The MOSAICS model adds a subgrid component of star cluster formation and evolution into the EAGLE simulations. When a stellar particle forms in the simulations it has the ability to convert some of its (subgrid) mass into a star cluster population. The star clusters then acquire the position, velocity, age and chemistry of their host stellar particle. The fraction of mass assigned for star cluster formation in a stellar particle is determined by the cluster formation efficiency (CFE) and is dependent on the local natal gas pressure \citep{Kruijssen2012}. Once the fraction of mass has been assigned, cluster masses are stochastically drawn from a \citet{Schechter1976} mass function with an exponential truncation mass $\Mc$ that is environmentally dependent \citep{Reina-Campos2017},
\begin{equation}
dN/dM \propto M^{\alpha}\exp{(-M/M_{c,*})},
\end{equation}
where $\alpha$ is the index of the power law part of the function for masses below the truncation mass $\Mc$, above which the function can be described by an exponential. 
Observations of YMC populations indicate that $\alpha \simeq -2$ (e.g. \citealt{Zhang1999,McCrady2007,Dowell2008,Baumgardt2013,Adamo2018}). The giant molecular cloud mass function has a similar, but slightly shallower slope (see \citealt{Chevance2020} for a recent review). This power law index has been used to describe the distributions of masses of many stellar systems within a galaxy and is likely the result of fragmentation produced by the balance between gravitational collapse and turbulence compression \citep{Elmegreen2011,Adamo2020}.\par

The \citet{Reina-Campos2017} model assumes that $\Mc$ is proportional to the mass of the most massive molecular cloud \citep{Kruijssen2014}. The model predicts that the largest cloud mass is set by the interplay between the gravitational collapse of the largest unstable region in a differentially-rotating disk, i.e. the \citet{Toomre1964} mass, and stellar feedback from the newborn stellar population within that region. If stellar feedback halts and disperses the collapsing region, the resulting cloud mass is smaller than the \citet{Toomre1964} mass. This model predicts that, in the feedback-limited regime, $M_{\rm c,*}$ increases in environments with higher gas pressures, as feedback becomes less efficient in such environments.\par 

Star clusters lose mass in the simulations through stellar evolution and dynamical processes. Stellar mass-loss follows the EAGLE model \citep{Wiersma2009}. Dynamical mass-loss is due to two-body relaxation, tidal shocks and total disruption through dynamical friction. The contributions from two-body relaxation and tidal shocks are calculated via the local tidal tensor and dynamical friction is applied at every snapshot in post-processing. The dynamical friction timescale is calculated for all clusters at each simulation snapshot and clusters can be completely removed by dynamical friction when the dynamical friction timescale is less than the age of the cluster \citep[see][for full details]{Pfeffer2018}.\par

\citet{Pfeffer2019YC} compare the $\Mc$ of young star cluster populations in the E-MOSAICS simulations with observations of local galaxies and find good agreement, though more observations of systems with young star clusters are needed to test whether the scatter found in the simulations is realistic. In this work we expand on the \citet{Pfeffer2019YC} study to contrast the $z=0$ $\Mc$ of GC systems (with no age constraints) to that of observations. This means that we allow the E-MOSAICS initial cluster mass function to evolve with time (through stellar evolution and dynamical processes) and test whether observations are still matched. This is a test of the cluster formation physics in the simulations and also the subsequent evolution of the star clusters alongside their host galaxies.

\section{Comparison between simulations and observations \label{3}}
\subsection{Virgo Cluster Data}
The data we compare our simulations to throughout this work are part of the ACS Virgo Cluster Survey, first presented by  \citet{Cote2004}. The survey is designed to observe 100 early-type galaxies in the Virgo Cluster, using the Advanced Camera for Surveys (ACS) on the Hubble Space Telescope. The survey used the F475W and the F850LP bandpasses (approximately equal to the Sloan g and z respectively, \citealt{Cote2004}). The ACS Virgo Cluster Survey is designed to probe the brightest $\approx 90 \%$ of the GC luminosity function in the 100 galaxies. This yields a sample of $\approx 13,000$ GCs in the Virgo Cluster \citep{Cote2004}. 

\citet{Jordan2007} present the luminosities of GCs belonging to early-type galaxies in the ACS Virgo Cluster Survey. They fit the luminosity functions with an evolved Schechter function (which is meant to account for the GC mass loss) and present the truncation luminosity in their table 3, and the corresponding truncation mass ($\Mc$) as a function of galaxy stellar mass based on the B-band galaxy magnitude in their figure 16. \par

\citet{Peng2008} present the stellar masses of the galaxies in the ACS Virgo Cluster Survey. We use their table 1 to obtain the stellar mass for each of the galaxies in the \citet{Jordan2007} sample. The \citet{Jordan2007} results in this format are presented as black stars in Fig. \ref{fig:galmass_fiducial}.

\subsection{Simulation Data}

To compare the mass functions of GCs in the E-MOSAICS simulations with those in the Virgo galaxy cluster \citep{Jordan2007}, we use the most massive galaxy group in the simulation, which has a virial mass $M_{\mathrm{200}} = 5.14 \times 10^{13} \Msun$\footnote{The total mass contained within the radius at which the density drops to 200 times the critical density.}. The stellar mass of the brightest cluster galaxy (BCG) is $M_{*}=2.23 \times 10^{11} \Msun$ and the cluster contains 154 galaxies with a stellar mass above $10^{7} \Msun$. The virial mass of the Virgo galaxy cluster has been estimated to be $6.3 \times 10^{14} \Msun$ by \citet{Kashibadze2020}  and $4.2 \times 10^{14} \Msun$ by \citet{McLaughlin1999}. 
This places our simulated galaxy cluster at a lower mass than the Virgo galaxy cluster but similar to the Fornax cluster \citep{Drinkwater2001}.
\citet{Villegas2010} showed that the GC mass function dispersion and turnover mass is similar in both the Virgo and Fornax clusters, while \citet{Liu2019} showed the behaviour of specific frequency with galaxy luminosity is also similar. 
Therefore we consider the comparison of GCs in simulated Fornax-mass clusters with the Virgo cluster to be reasonable.
We also show in Section \ref{3.3} that halo mass does not have a strong impact on our results.

\subsubsection{Description of the Schechter function fits}

To fit Schechter functions to the GC mass functions from the simulations, we follow the methodology outlined by \citet{Pfeffer2019YC}, who adopt similar analyses to those used in observational studies (e.g. \citealt{Johnson2017,Messa2018a}). We stack all the GCs in the simulated cluster in bins of host galaxy stellar mass and use the Markov Chain Monte Carlo (MCMC) code PyMC \citep{Fonnesbeck2015} to perform the fits and to sample the posterior distribution of the Schechter power-law index and truncation mass. Stacking galaxies is necessary to increase GC numbers for robust Schechter fits. \citet{Jordan2007} also stack galaxies in the lower-mass galaxy mass bins. The power law index is sampled with a uniform prior between -3 and -0.5. The truncation mass is sampled in log-space with a uniform prior between a minimum cluster mass (which we describe below) and $10^9 \Msun$. We use a Gaussian likelihood for $\log\Mc$. For each fit the MCMC takes 10,000 steps and we take the first 1,000 of these steps as burn-in.\par 

In the galaxy mass range of the \citet{Jordan2007} observations ($\log{M_*/M_{\odot}} > 9.5$) we bin the galaxies by stellar mass in bins of width $0.25$ dex. Below this mass, we use galaxy stellar mass bins of width 0.5 dex to yield the best sampling. We present the number of galaxies and number of GCs used in Table \ref{table:GCnumber}, where the `fiducial', `No dynamical friction (DF)' and `initial' columns refer to three GC subsamples from the simulations that are described in Section \ref{5}. In the range of the observations ($\log{M_*/M_{\odot}} > 9.5$) the simulated sample contains a similar number of galaxies as the \citet{Jordan2007} study, who also include just 1 galaxy in their most massive bin and 10 in their least massive. We do, however include more GCs in each bin than \citet{Jordan2007}: their sample spans a range of 193-1721 GCs, whereas our fiducial sample of GCs spans a number range of 612-2992 within the same galaxy mass range. Due to the lack of a cold, dense gas phase in the EAGLE model, it is a known problem that E-MOSAICS does not disrupt enough low-mass GCs \citep[see][]{Pfeffer2018, Kruijssen2019a}. Since the $\Mc$ relies on a fit to the high-mass end of the GC mass function this should not affect our $\Mc$ results, but could have an impact on the reliability of the fit to the slope.
Indeed, for this reason, we do not attempt to fit turnover masses to the mass functions.\par

We use a varying minimum cluster mass across the galaxy mass range, in order to restrict the fits to the high-mass end of the GC mass functions (however in Appendix~\ref{A} we consider the effect of different minimum masses for the fits).
We therefore fit Schechter functions to the upper 2 dex of the mass function, from the third most massive GC to account for stochasticity at the high-mass end.
We find this range sufficient to enable fitting of both the upper truncation ($\Mc$) and power-law index ($\alpha$).
We quote the minimum masses of the GCs in Table \ref{table:GCnumber}.
The minimum mass adopted also scales with galaxy mass in a similar way to the completeness limit for Virgo GCs. For example, from table 1 in \citet{Jordan2007}, Virgo galaxies with stellar masses $M_\ast \approx 10^9 \Msun$ have a GC completeness of 90 per cent at the luminosity limit $m_z = 25.2$, corresponding to a GC mass of $\approx 2 \times 10^4 \Msun$. 
Virgo galaxies with mass $M_\ast \approx 10^{11} \Msun$ reach GC completeness of 90 per cent at $m_z \approx 23$, corresponding to a GC mass of $\approx 1.5 \times 10^5 \Msun$.
Both values are similar to the minimum masses we adopt.
We note that the lower galaxy mass bins, outside of the galaxy mass range of the \citet{Jordan2007} observations, include GCs that would be too faint to observe at the distance of the Virgo cluster, but it is still interesting from a theoretical stand point to investigate the continuation of the trend at lower galaxy stellar masses.


\subsubsection{Should we fit upper truncations?}
There is some contention in the literature as to whether Schechter functions or single power laws fit star cluster mass functions more accurately \citep{Chandar2014,Chandar2016,Mok2019}. Although on a per-particle basis the mass function assumed by the E-MOSAICS fiducial model is a Schechter function with an environmentally dependent $\Mc$, this does not necessarily mean that the final mass function will be best fit by a Schechter function. The GC mass function of each simulated galaxy is an accumulation of the GC populations associated with many particles; each with varying input $\Mc$, and dynamical evolution may erase the signal of any exponential truncation. \par

Therefore, we also fit power law functions to the GCs in each galaxy mass bin, over the same mass range and with the same MCMC method as described for the Schechter fits above. The power law index is sampled with a uniform prior between -3 and -0.5. We then calculate the Bayesian Information Criterion (BIC) value \citep{Schwarz1978} for both of the fitting functions and compare them.\par

The BIC value takes into account that a model with more free parameters is likely to fit the data better and penalises the maximum likelihood estimate of the model if there are more free parameters in the fit. The BIC value is given by, 
\begin{equation}
\mathrm{BIC} = k \ln{(n)} - 2\ln{(L)}
\end{equation}where $k$ is the number of free parameters in the fit, $n$ is the sample size, $L$ is the maximum likelihood estimate of the model and lower BIC values are favourable. We can compare two models by calculating the difference in their BIC values ($\Delta$BIC). When we subtract the BIC value of the Schechter fit from that of the power law fit, positive values indicate that the Schechter function is preferred over the power law and vice versa. We find that, for all galaxy mass bins, a Schechter function is strongly preferred with $\Delta$ BIC values between 18 and 133. In Section \ref{5}, we include two more subsamples of GCs which omit specific mass loss mechanisms. A Schechter function is also preferred over a power law function for all galaxy masses in these subsamples. For the subsample of GCs with dynamical friction omitted, $\Delta$ BIC values are between $12 - 105$, and for the subsample of GCs with all mass loss omitted, $\Delta$ BIC values are between $ 2 - 129$. Given that all $\Delta$ BIC values are positive we are confident that the simulated globular cluster mass functions are better fit by a function with an upper truncation mass.\par

We compare the effect of varying minimum masses and also the results from the pure power-law fit in Appendix~\ref{A}.

\begin{center}
\begin{table}
 \caption{\label{table:GCnumber} The number of galaxies and GCs in each galaxy mass bin and GC sub-population. The highest galaxy mass bins contain just one galaxy, so in this case we give the galaxy's mass.}
 \resizebox{\hsize}{!}{
\begin{tabular}{ |c|c |c|c|c|c| } 
 \hline
 $\log{(M_{*}/\Msun)}$ & Galaxies&Min. GC mass [$\Msun$]&Fiducial& No DF & Initial  \\
 \hline
7--7.5 &41&$1.63 \times 10^{3}$&441&476&2609\\
7.5--8&39&$2.66 \times 10^{3}$&884&913&6236\\
8--8.5 &18&$3.24 \times 10^{3}$&1184&1211&10690\\
8.5--9 &15&$6.67 \times 10^{3}$&1877&1898&17958\\
9--9.5 &10&$1.60 \times 10^{4}$&1901&1923&14642\\
9.5--9.75 &6&$2.70 \times 10^{4}$&2494&2522&16624\\
9.75--10 &4&$5.45\times 10^{4}$&1301&1844&21364\\
10--10.25 &7&$5.68 \times 10^{4}$&2465&3137&55148\\
10.25--10.5 &2&$4.21 \times 10^{4}$&1484&1618&33617\\
10.67 &1&$5.33 \times 10^{4}$&612&719&36666\\
10.91 &1&$7.16 \times 10^{4}$&1323&1548&47650\\
11.05 &1&$1.18 \times 10^{5}$&1410&1801&45898\\
11.35 &1&$1.14 \times 10^{5}$&2992&3332&42234\\
 \hline
 \end{tabular}}
 \end{table}
\end{center}

\subsection{Truncation masses in simulations vs. Observations}\label{3.3}

\begin{figure}
    \centering
	\includegraphics[width=\linewidth]{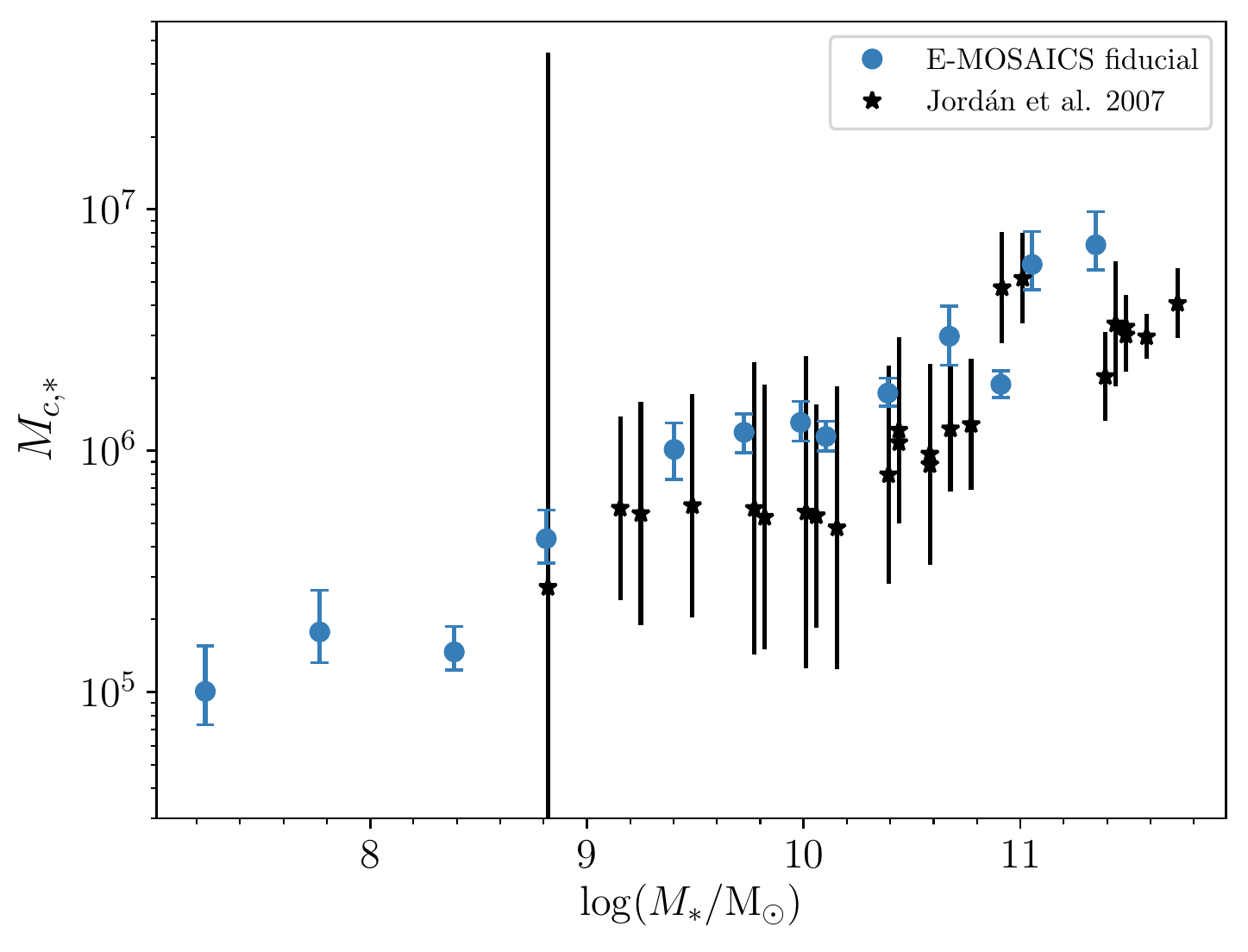}
    \caption{The dependence of $\Mc$ on galaxy stellar mass in the Virgo galaxy cluster and the most-massive E-MOSAICS galaxy group. The black stars represent the data taken from Fig. 16 of \citet{Jordan2007}. The blue points show the fits to the E-MOSAICS fiducial model at z=0, where the error bar represents the 16th-84th percentile range of the posterior distribution. The blue E-MOSAICS points match well with the \citet{Jordan2007} sample. }
    \label{fig:galmass_fiducial}
\end{figure}

\begin{figure}
    \centering
	\includegraphics[width=\linewidth]{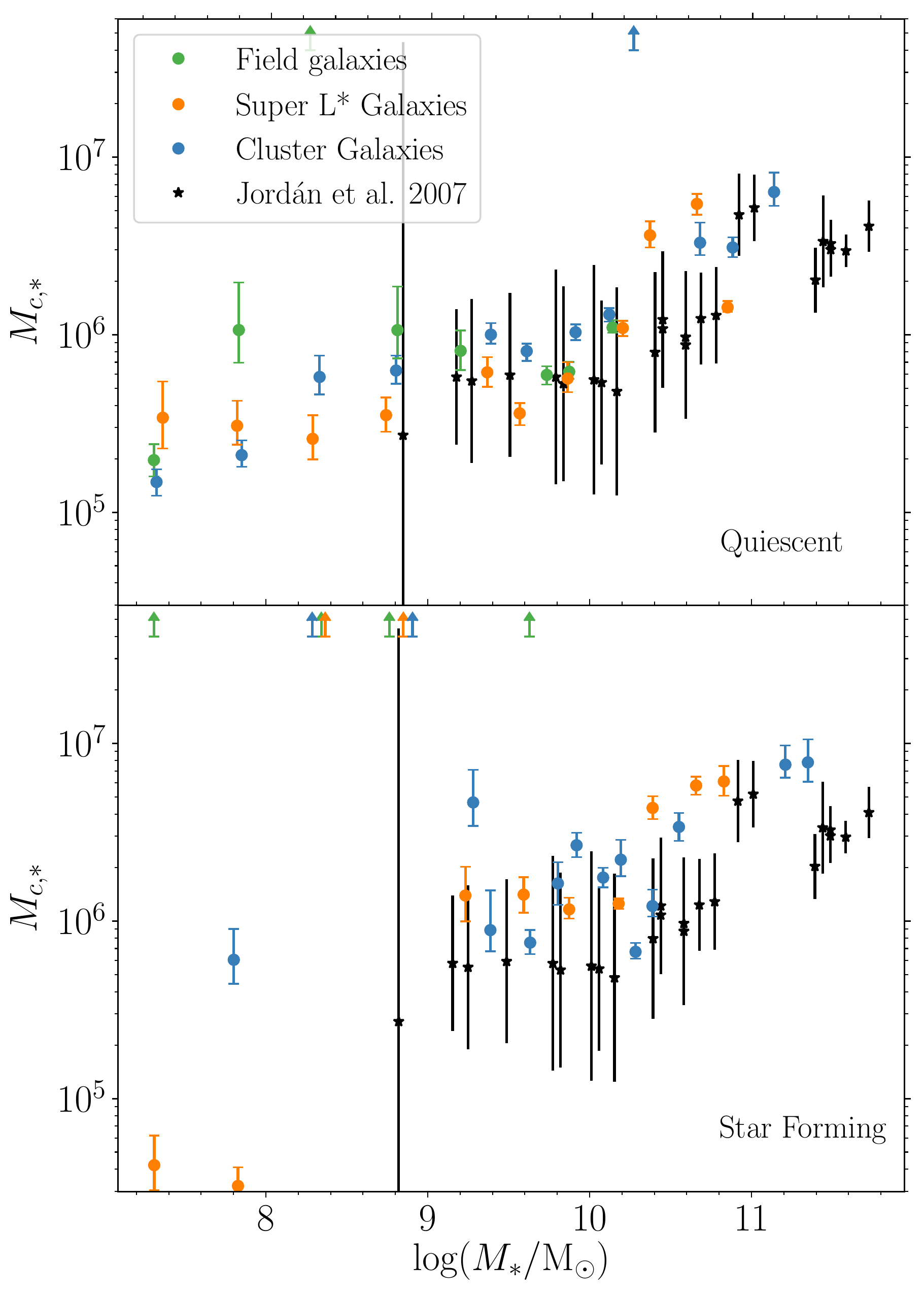}
    \caption{The dependence of $\Mc$ on galaxy stellar mass in the Virgo galaxy cluster and the most-massive E-MOSAICS galaxy group. The black stars represent the data taken from Fig. 16 of \citet{Jordan2007}. The coloured points show the fits to the E-MOSAICS volume, split by the group mass `field': $M_{200} < 10^{12} \Msun$; `super L$^\ast$': $10^{12} < M_{200} / \mathrm{M}_{\sun} < 10^{13}$; cluster: $M_{200} > 10^{13} \Msun$). The top panel shows the quiescent galaxies and the bottom panel shows star forming galaxies.}
    \label{fig:galmass_fiducial_volume}
\end{figure}

In Fig. \ref{fig:galmass_fiducial} we show $\Mc$ as a function of galaxy stellar mass for the observations from \citet{Jordan2007} and for the most massive galaxy group in the E-MOSAICS simulations. Error bars show the 16th and 84th percentiles of the posterior distribution. Fig. \ref{fig:galmass_fiducial} shows good agreement between the truncation masses of the observations and the E-MOSAICS simulations, particularly in the slope of the relation. There is a systematic offset between the two distributions in Fig.  \ref{fig:galmass_fiducial}: $\Mc$ is consistently higher in E-MOSAICS than in the observations. This offset is smaller than the observational uncertainties, but could be due a number of effects. On the simulation side, under-disruption could mean that $\Mc$ is overestimated (see Appendix D of \citealt{Kruijssen2019a}), the dynamical friction timescales (calculated in post-processing) could be too long, or the initial $\Mc$ slightly too large. Alternatively, some of the difference may be due to uncertainties in colour-$M/L$ conversions for observed GCs. \par

In Fig. \ref{fig:galmass_fiducial_volume} we show the same observational data from \citet{Jordan2007} as in Fig. \ref{fig:galmass_fiducial} but we now show $\Mc$ as a function of galaxy mass for all galaxies in the E-MOSAICS volume, divided by whether they are quiscent or star forming and their group mass. Upward arrows represent the galaxy mass bins where a robust Schechter fit could not be acheived. We define `field galaxies', `super L* galaxies' and `cluster galaxies' as having total group masses $M_{200}<10^{12}$, $10^{12}<M_{200}/\mathrm{M}_{\sun}<10^{13}$ and $M_{200}>10^{13} \Msun$ respectively. Following the \citet{Pfeffer2019YC} method, we select galaxies that are currently not on the star forming main sequence based on the specific star formation rate (sSFR) within a $30 \kpc$ aperture. The reasoning for splitting galaxies by their current star formation is two-fold: firstly, galaxies that reside in galaxy clusters (such as those used in \citealt{Jordan2007}) are likely to be quiescent and secondly, galaxies that are forming massive GCs at $z=0$ could bias the $\Mc$ fits to high values. Fig. \ref{fig:galmass_fiducial_volume} shows that quiescent galaxies in the full E-MOSAICS volume also show agreement in the truncation masses with the \citet{Jordan2007} observed values, whereas those that are star forming have, on average, slightly higher $\Mc$ (possibly due to having slightly younger, less evolved GCs). In the star forming group some galaxy mass bins could not get a Schechter fit, likely due to greater stochasticity of high-mass GCs. This confirms that there is nothing special about the galaxies in the most massive galaxy group except that they are likely to be quiescent. Therefore we continue the rest of this work with the most massive galaxy group to ease comparison with observations and to simplify discussion.

Note that in the model, the $\Mc$ is calculated separately to the power law index. Therefore if the power law index was shallower/steeper, there would be more/fewer GCs at higher masses and although $\Mc$ would be better/worse sampled, its value would remain the same.

Given the above discussion it is still fair to suggest that the E-MOSAICS simulations show mass function truncations that are a satisfactory match to the \citet{Jordan2007} observations and make the clear prediction that $\Mc$ increases with galaxy mass. This demonstrates that the fiducial input physics of the MOSAICS model is able to reproduce a fundamental observable in GC studies. To determine which physical mechanism is the most important in setting the relation in Fig \ref{fig:galmass_fiducial}, we will examine alternative formation physics in the next section.

\section{Alternative cluster formation physics \label{4}}
\begin{figure}
    \centering
	\includegraphics[width=\linewidth]{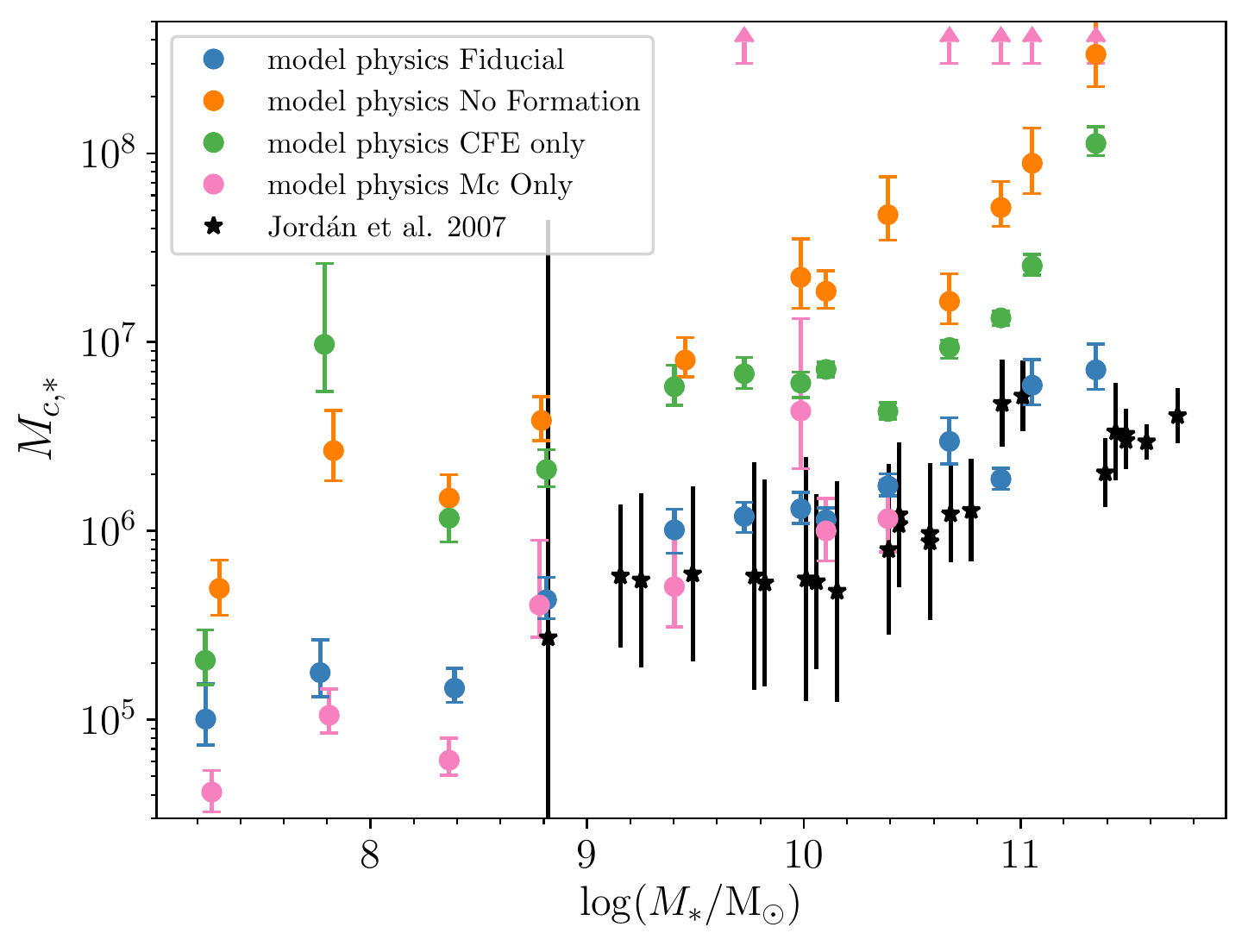}
    \caption{The dependence of $\Mc$ on galaxy stellar mass in the Virgo galaxy cluster and the most-massive E-MOSAICS galaxy group. The fiducial model, the `no formation' model, the `CFE only' model and the `Mc only' model are represented by blue, orange, green and pink circles respectively.}
    \label{fig:Mcstar_galmass_fiducial_altphys.pdf}
\end{figure}

In this section we investigate three alternative cluster formation physics variants in the E-MOSAICS model to establish which of the key ingredients of the model are needed to reproduce the observations of \citet{Jordan2007}. We outline the differences between the models below:
\begin{itemize}
\item In the fiducial model, both the cluster formation efficiency and $\Mc$ depend on environment (as described in Section \ref{2})
\item In the `no formation physics' model, there is a constant cluster formation efficiency ($\Gamma = 0.1$) and no upper truncation to the mass function i.e. it is a pure power law, therefore the cluster formation is not environmentally dependent and is equivalent to a simple "particle tagging" method.
\item In the CFE only model, the CFE varies with environment but there is no upper truncation to the mass function.
\item In the $\Mc$ only model the CFE is a constant ($\Gamma = 0.1$) but $\Mc$ varies with environment.
\end{itemize}

In Fig. \ref{fig:Mcstar_galmass_fiducial_altphys.pdf} we present the four alternative physics models in blue, orange, green and pink respectively. Firstly we will focus on the `no formation physics' model (orange), where the increasing trend of $\Mc$ with galaxy stellar mass is simply a size-of-sample effect. More massive galaxies form more GCs and therefore have the potential to sample more massive GCs from the power-law mass function. Dynamical friction then acts to remove some of the most massive GCs and a truncation is detected. The slope of the relation will be constant, but the relation could be shifted up or down, depending on the CFE. However, the slope of the `no formation physics' model is significantly steeper than that of the observations, with slopes of 0.58 and 0.40 respectively, so even with a smaller CFE to shift the relation to lower $\Mc$, it would not match the observations.\par

Next, we concentrate on the `CFE only' model, where the increasing trend of $\Mc$ with galaxy stellar mass is still present and mostly follows that of the `no formation model' except for a dip in $\Mc$ at $\log(M_{*}/M_{\odot}) \approx 10$. The dip occurs because there is now an environmentally dependent CFE so galaxies forming most of their GCs in high pressure regions have the potential to form their most massive GCs here as well, in an environment that can subsequently disrupt them. Therefore it is likely that dynamical evolution is the cause of this slight decrease in $\Mc$, which we discuss further in the context of the fiducial model in Section \ref{5}.\par

Finally, we turn our attention to the `$\Mc$ only' model, here we mark with upward arrows the galaxy mass bins where a robust Schechter fit could not be achieved. We also carry out a BIC test for all the fits in Fig. \ref{fig:Mcstar_galmass_fiducial_altphys.pdf} to indicate whether a Schechter fit or a power law fit is more appropriate for the data. The BIC tests for the fits here indicate that a Schechter function is preferred in all cases except for those that are shown with an upward arrow, where a power-law fit is strongly preferred. It is interesting that in the case where a truncation mass is explicitly included in the model, a fit that does not include one is preferable in some galaxy mass bins. In the `fiducial' model both $\Mc$ and CFE scale with birth pressure, therefore where the $\Mc$ is high, also a higher fraction of the mass of the stellar particle is available for GC formation. By contrast in the `$\Mc$ only' model $\Mc$ scales with birth pressure and the CFE does not. As a result, there is less mass available and stellar particles are less likely to form massive GCs. In the `fiducial' model, high $\Mc$ particles contribute more clusters to the composite cluster mass function than low $\Mc$ particles due to the varying CFE but in the $\Mc$ only model all particles are weighted equally. Therefore when many particles are stacked in the mass function, the power-law index ($\alpha$) becomes steeper and an $\Mc$ is difficult to identify. \par

Together, the results in this section confirm that an environmentally varying CFE and mass function truncation, as implemented in the `fiducial' E-MOSAICS model, is required to explain both the GC and young cluster populations \citep{Pfeffer2018,Pfeffer2019YC,Usher2018,ReinaCampos2019,Bastian2020}.

\section{Dependence on GC dynamical evolution \label{5}}

\subsection{GC mass loss models}

As described in Section \ref{2}, the main GC mass loss mechanisms are stellar evolutionary mass loss, tidal shock heating and two body relaxation. Clusters can be completely removed via dynamical friction. Here we investigate the different mass loss mechanisms and how they affect the GC mass function. We again fit Schechter functions to the GCs in the same galaxy mass bins as in Fig. \ref{fig:galmass_fiducial} but we now include two new subsamples of GCs. We include the GCs from the simulations without dynamical friction applied, shown in orange in Figs. \ref{fig:Mcstar_galmass_massloss.pdf} - \ref{fig:alpha_galmass_massloss.pdf}, and also the initial GCs that formed with a mass greater than the minimum mass given in Table \ref{table:GCnumber}, with no mass loss -- stellar or dynamical-- applied, shown in green in Figs. \ref{fig:Mcstar_galmass_massloss.pdf} - \ref{fig:alpha_galmass_massloss.pdf}. We note that stellar mass loss should affect all GCs approximately equally (assuming the GCs are relatively old) and therefore will not play a part in setting the slope of the GC mass function.\par

\subsubsection{The truncation mass}
\begin{figure}
    \centering
	\includegraphics[width=\linewidth]{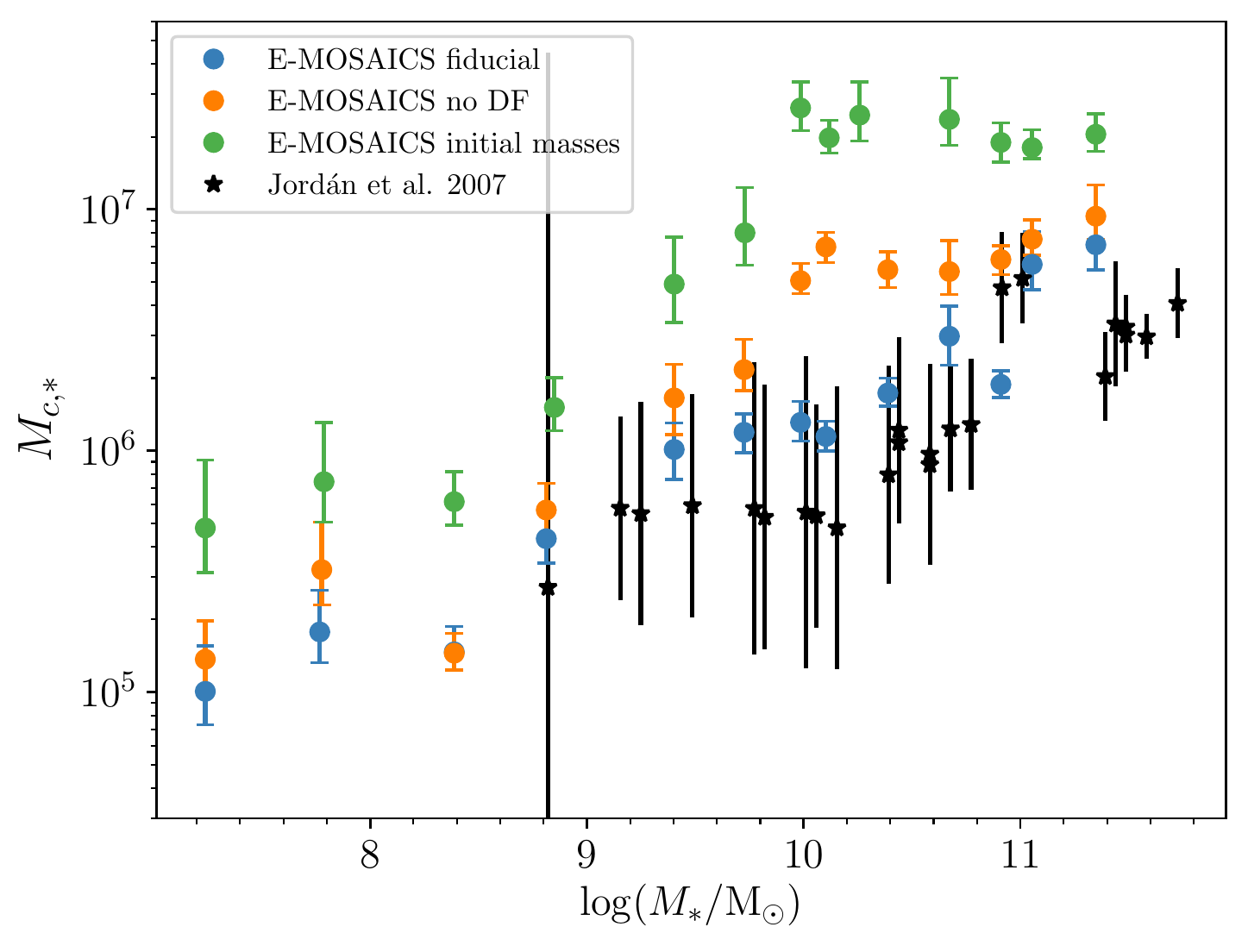}
    \caption{The dependence of $\Mc$ on galaxy stellar mass in the Virgo galaxy cluster and the E-MOSAICS galaxy cluster. The black stars represent the data taken from Fig. 16 of \citet{Jordan2007}. The blue points show the E-MOSAICS fiducial model at z=0. The orange points show the E-MOSAICS model with no dynamical friction taken into account. Finally, the green points show the E-MOSAICS with no mass loss (stellar evolution or dynamical) taken into account. All dynamical evolution processes must be included in the simulation to match well with the \citet{Jordan2007} sample. }
    \label{fig:Mcstar_galmass_massloss.pdf}
\end{figure}

\begin{figure}
    \centering
	\includegraphics[width=\linewidth]{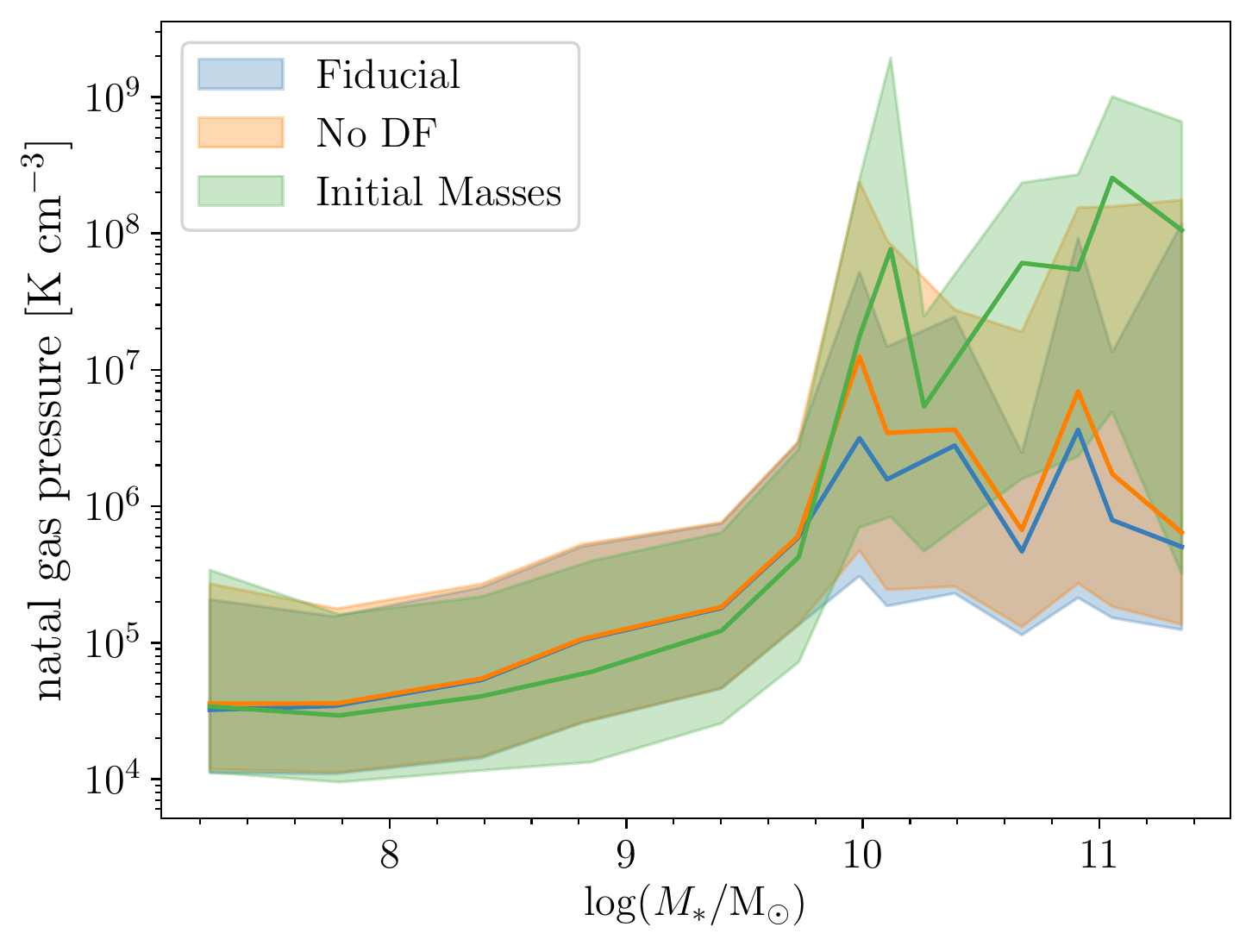}
    \caption{The dependence of GC birth pressure on galaxy stellar mass in the E-MOSAICS galaxy cluster. The shaded regions show the 16th-84th percentile range.}
    \label{fig:galmass_birthpressure.pdf}
\end{figure}

\begin{figure}
    \centering
	\includegraphics[width=\linewidth]{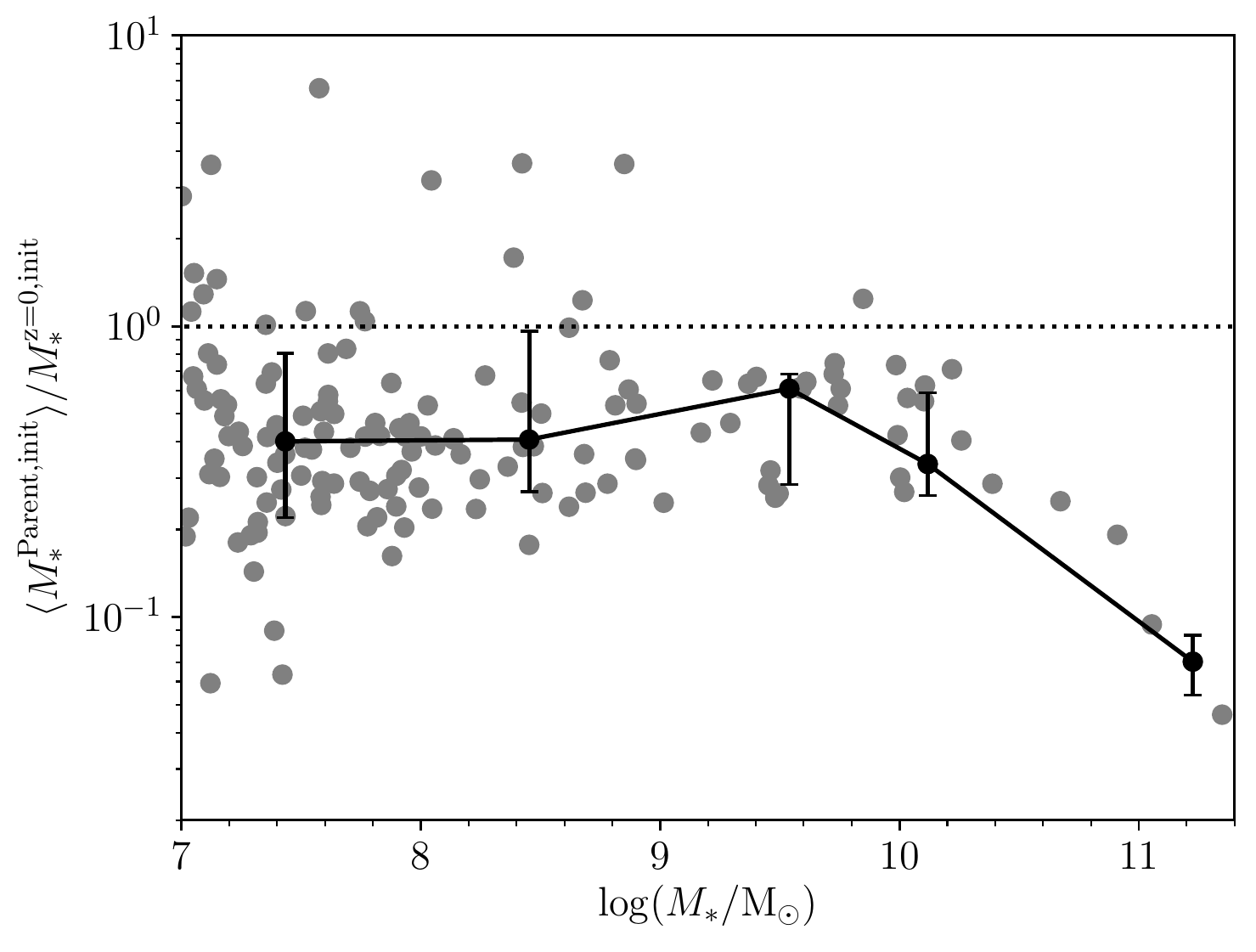}
    \caption{The ratio of the median initial parent galaxy stellar mass to the initial stellar mass of each GC-hosting galaxy in the E-MOSAICS galaxy cluster at $z=0$. The black points and line show the median and 16th-84th percentile ranges in galaxy mass bins of 1 dex.}
    \label{fig:mass_mass.pdf}
\end{figure}

\begin{figure}
    \centering
	\includegraphics[width=\linewidth]{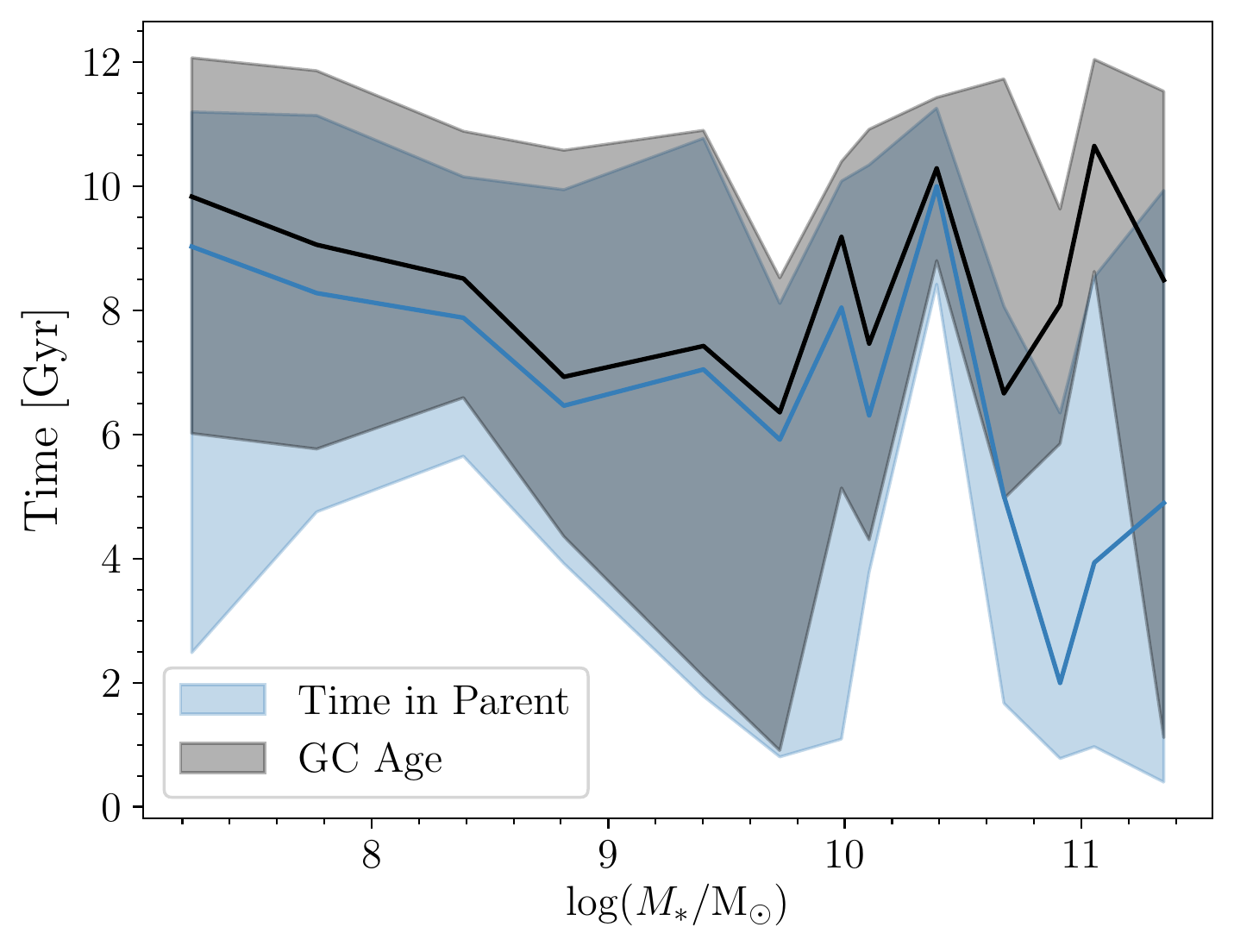}
    \caption{The dependence on the median time a GC spends in its parent galaxy as a function of galaxy stellar mass. The colours of the three subsamples are the same as in Fig. \ref{fig:Mcstar_galmass_massloss.pdf} and the shaded regions show the 16th-84th percentile range. }
    \label{fig:galmass_TimeInParent.pdf}
\end{figure}

In Fig. \ref{fig:Mcstar_galmass_massloss.pdf} we show the fitted $\Mc$ as a function of galaxy stellar mass, with our fiducial, no dynamical friction (no DF) and no mass loss (initial masses) GC models shown in blue, orange and green respectively, and the \citet{Jordan2007} fits shown as black stars. Fig. \ref{fig:Mcstar_galmass_massloss.pdf} shows that the subsample of GCs that does not include any mass loss (initial masses) has the highest $\Mc$, followed by the subsample that includes all mass-loss mechanisms except dynamical friction (no DF) and then the fiducial E-MOSAICS model exhibiting the lowest $\Mc$ in a given galaxy mass bin. The $\Mc$ decreases by $\sim 40$ per cent due to stellar evolution, however any further decrease is due to dynamical evolution. This indicates that the GC disruption time-scale is short enough to destroy high mass GCs, and that dynamical evolution plays an important role in shaping the high mass end of the GC mass function. \par

We relate the galaxy mass to the birth pressure of the GCs in Fig. \ref{fig:galmass_birthpressure.pdf}, which shows the birth pressure of the different subsamples of GCs in the same galaxy mass bins as in Fig. \ref{fig:Mcstar_galmass_massloss.pdf}, where the solid line shows the median and the shaded region represents the 16th-84th percentile range. The birth pressures of the three subsamples are very similar up to a galaxy mass of $ \approx 10^{10} M_{\odot}$, because the samples themselves are similar, i.e. there are not many GCs formed that do not survive until present day. Above a galaxy mass of $\approx 10^{10} M_{\odot}$ there is a steep increase in the initial birth pressures before a plateau. It is also at this mass where there is a separation in the median birth pressures between the `initial masses' sample ($P/k \sim 10^8 \, \mathrm{K} \, \mathrm{cm}^{-3}$) and the other two populations ($P/k \sim 10^6 \, \mathrm{K} \, \mathrm{cm}^{-3}$). \par

This separation occurs because of the high birth pressures of the initial GCs. In high pressure/high density environments, mass loss mechanisms such as tidal shocks are more prevalent and therefore quickly disrupt the newly formed GCs (termed the `cruel cradle effect', \citealt{Kruijssen2012CC}; see also Section 6.2 of \citealt{Pfeffer2018} in relation to the E-MOSAICS simulations). The tidal disruption timescale is much smaller than the dynamical friction timescale; therefore, before dynamical friction can act, GCs that have formed in the highest pressure environments have already been disrupted. Dynamical friction then becomes effective at reducing $\Mc$ at a galaxy mass of $\approx 10^{10} M_{\odot}$ but this is not reflected in the birth pressures. This is simply because dynamical friction removes the most massive GCs that are few in number, so although this will affect the $\Mc$ it will not affect the median birth pressures of the surviving GCs.\par

An interesting feature of Figs. \ref{fig:Mcstar_galmass_massloss.pdf} and \ref{fig:galmass_birthpressure.pdf} is that both the $\Mc$ of the initial GCs and the natal birth pressure show a plateau above $\log(M_{*}/M_{\odot}) \approx 10$ while we might intuitively expect a continuing increase with galaxy mass, with more massive galaxies able to form a greater number of more massive GCs. We must consider, however that massive galaxies grow via mergers and therefore the massive galaxy we observe at $z=0$ is an accumulation of many galaxy building blocks. Therefore, we must investigate not the galaxy mass at $z=0$ but rather the galaxy mass at the time of GC formation. For this, we compare the median stellar mass of the galaxies in which the `initial' GCs formed (the parent galaxy mass) to the $z=0$ galaxy stellar mass. This is shown in Fig. \ref{fig:mass_mass.pdf}, where each grey point represents one galaxy in the cluster and the black line shows the median for galaxy mass bins of 1 dex.  Note that the y-axis shows the ratio of the stellar mass without stellar evolution taken into account, this is to remove the effect of some galaxies having more evolved stellar populations. Fig. \ref{fig:mass_mass.pdf} shows that the median GC parent galaxy mass relative to the $z=0$ galaxy mass is broadly constant for $\log(M_{*}/M_{\odot}) < 10$, but it declines to higher masses. Above a stellar mass of $10^{10} \Msun$ there is a higher fraction of GCs that were born in a lower mass galaxy compared to their host galaxy at $z=0$. Therefore, the birth pressures and subsequently $\Mc$ remain constant, even with increasing $z=0$ mass. This is because massive galaxies increasingly grow by mergers, not star formation, so they are unlikely to be forming new GCs during their late accretion-driven growth stage (e.g. \citealt{Oser2010,Lee2013,Qu2017,Clauwens2018,Davison2020}) \par

Another interesting feature of Fig. \ref{fig:Mcstar_galmass_massloss.pdf} is that the effect of dynamical evolution is not constant across all galaxy mass bins. Dynamical mass loss has the most power at reducing $\Mc$ from the initial masses of all GCs to the masses at $z=0$ at $\log(M_{*}/M_{\odot})  \approx 10$.  As discussed previously, it is the addition of dynamical friction that drives the decrease in $\Mc$ at these galaxy masses through the removal of the most massive GCs. Here we discuss why this is more efficient at a galaxy mass of $\log(M_{*}/M_{\odot})  \approx 10$ than for $\log(M_{*}/M_{\odot})  \approx 11$.\par

The first contributing factor to longer dynamical friction timescales (and therefore a higher chance of survival) is the mass ratio between the GC mass and the galaxy mass. When the mass within the GC's orbit is larger, the dynamical friction timescale is longer ($\tau_\mathrm{df} \propto V_c$, \citealt{LaceyCole93}). In the more massive galaxies, the mass within the GC's orbit is likely to be larger at fixed radius and therefore the GC can survive for longer. The second contributing factor to longer dynamical friction timescales is the radius of the GC orbit. Importantly GCs may get pushed to wider orbits via mergers \citep[e.g.][]{Kruijssen2011}. Mergers facilitate the means for GCs to move from their birth places (where dynamical friction timescales may be short) either by being kicked out of the inner parts of the galaxy or being deposited in the halo of a more massive galaxy (where dynamical friction timescales are very long). \citet{Qu2017}, along with \citet{Clauwens2018} and \citet{Davison2020}, showed that the EAGLE galaxies are built by mainly in-situ star formation up to a stellar mass $\approx 10^{10} M_{\odot}$. The ex-situ fraction then increases with stellar mass, and for galaxies that reach a stellar mass $\approx 10^{11} M_{\odot}$ approximately 50 per cent of their mass is built through mergers. 
Thus, plausibly, it is the lack of redistribution of massive GCs by mergers which leads to more effective dynamical friction in $\sim 10^{10} \Msun$ galaxies.

To quantify this, we now consider how long the GCs in each $z=0$ galaxy typically spend in their parent galaxy (i.e. the time between GC formation and $z=0$ for in situ GCs, or the time between formation and the merger of the host galaxy in the case of accreted GCs). This will inform us about whether the GC population is dominated by GCs that have survived in their parent galaxies for a long time or by GCs that have been deposited into the halo of the more massive galaxy after spending a short amount of time in their parent galaxy. We examine this in Fig. \ref{fig:galmass_TimeInParent.pdf} where we present the median age and 16-84 percentile range of the GCs and the median time and 16-84 percentile range the GCs spent in the parent galaxy \citep[analogous to figure D2 in][]{Kruijssen2019a}. The median age of the GCs remains old (> $7 \Gyr$) at all galaxy masses. Note the slight decline in age with increasing mass from $10^{7}$ towards $10^{10} \Msun$, this is because more massive galaxies are likely to have entered the potential well of the galaxy cluster more recently, and, when they do enter the potential well of the galaxy cluster, they can hold onto their star forming gas for longer than their lower mass counter parts (see e.g. \citealt{Gunn1972,Hughes2019} for more details). The time spent in the parent galaxy traces the age of the GCs closely upto a mass of $\log(M_{*}/M_{\odot})  \approx 10.5$ where the time spent in the parent galaxy decreases, whilst the median age still remains old. This reflects the fraction of GCs accreted from satellites into the halo of the galaxy where the dynamical timescale is long and massive GCs can survive.

In conclusion, we suggest that it is the combination of more massive GC formation and then the subsequent mass dependence of the galaxy merger histories and the effect of dynamical friction that leads to the fiducial trend between $\Mc$ and galaxy mass in both the \citet{Jordan2007} work and the E-MOSAICS simulations.

 \subsubsection{The mass function slope}

\begin{figure}
    \centering
	\includegraphics[width=\linewidth]{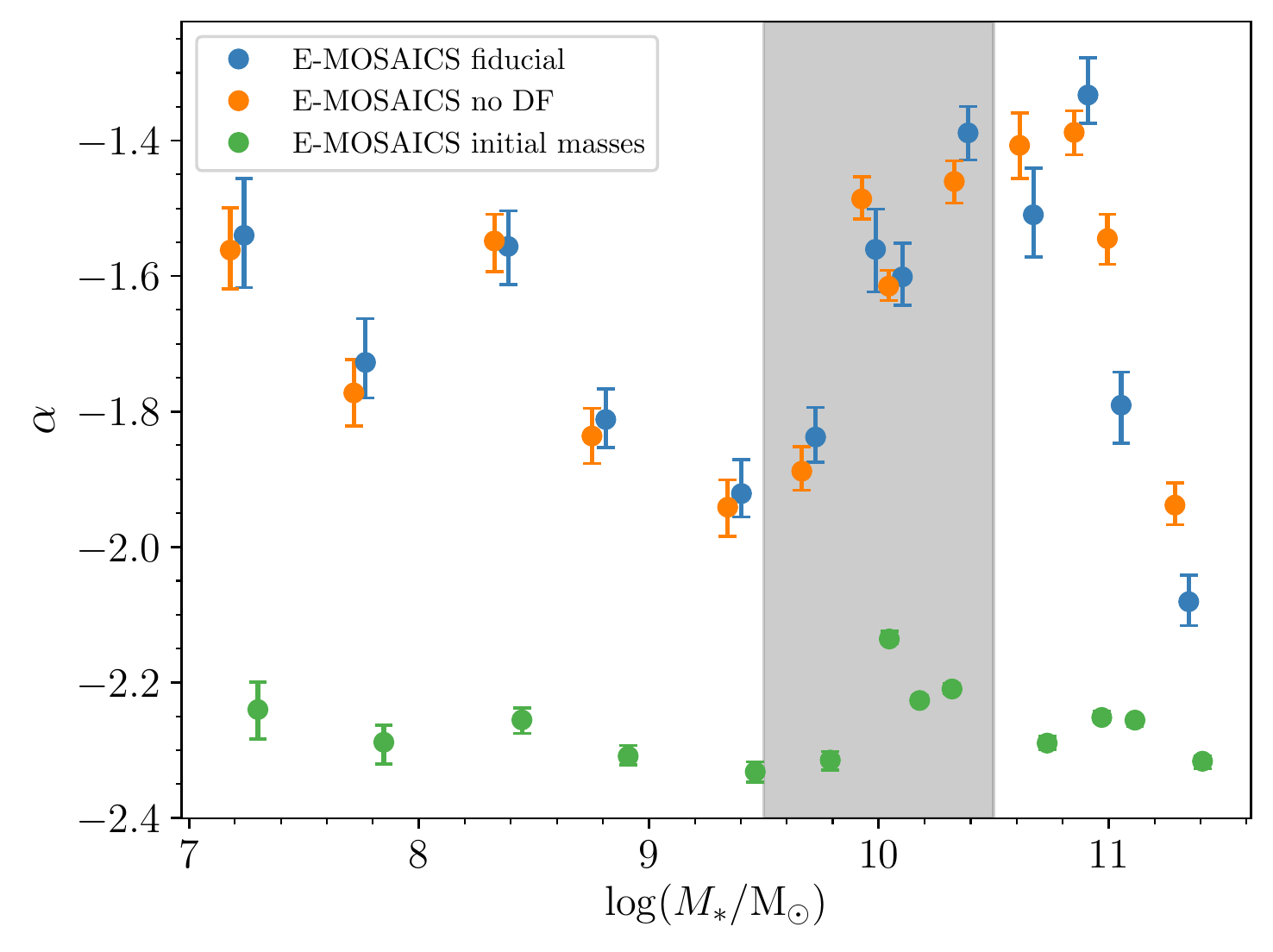}
    \caption{The dependence of $\alpha$ on galaxy stellar mass in the E-MOSAICS galaxy cluster. The blue points show the E-MOSAICS fiducial model at z=0. The orange points show the E-MOSAICS model with no dynamical friction taken into account. Finally, the green points show the E-MOSAICS clusters that survive to z=0 within our mass cut but with no mass loss (stellar evolution or dynamical) taken into account. The shaded region indicates where the increase of $\alpha$ with galaxy mass is due to increasing cluster mass loss.} 
    \label{fig:alpha_galmass_massloss.pdf}
\end{figure}

For completeness, we now consider the slope of the mass function in Fig. \ref{fig:alpha_galmass_massloss.pdf} where we show the power-law index ($\alpha$) of the Schechter fit in the same galaxy stellar mass bins, where the different coloured points have the same meaning as in Fig. \ref{fig:Mcstar_galmass_massloss.pdf}.
Due to the mass- and environmentally dependent cluster mass-loss mechanisms in the simulations (tidal shocks and two-body relaxation), the recovered power-law indices will depend on the minimum cluster mass for fitting and the strength of cluster mass loss in each galaxy.
We note that the minimum GC mass is different in each of these galaxy mass bins (see Table \ref{table:GCnumber}) because we are fitting the top two decades of the mass function. Therefore, the power-law index is derived from a different GC mass range in low-mass compared to high-mass galaxies. 
In particular, the minimum GC mass changes from $\sim 10^3$ to $\sim 10^4 \Msun$ for galaxy masses between $\log(M_{*}/M_{\odot}) \approx 7$-$9.5$, and from $\sim 10^4$ to $\sim 10^5 \Msun$ for galaxy masses $\log(M_{*}/M_{\odot}) \approx 10.5$-$11.5$.
Thus, some of the variation of $\alpha$ with galaxy mass will be caused by the varying minimum mass.

In Appendix~\ref{A} (lower panel of Fig.~\ref{fig:Mcstar_galmass_minmass.pdf}) we compare power-law indices with different assumptions for the minimum cluster mass in the mass function fits.
For masses $\log(M_{*}/M_{\odot}) < 9.5$ and $>10.5$ the power-law indices from the Schechter fits do not depend strongly on the adopted lower mass limit.
For masses $9.5 < \log(M_{*}/M_{\odot}) < 10.5$ (also highlighted in Fig.~\ref{fig:alpha_galmass_massloss.pdf}) there is a strong dependence on $\alpha$ with with minimum mass, such that the mass functions are steeper with increasing lower mass limit.
$\alpha$ increases with galaxy mass in this mass range, that is, the slope of the power-law portion of the mass function becomes shallower. We can associate this with a rise in the birth pressures with increasing galaxy mass across the same galaxy mass range (Fig.~\ref{fig:galmass_birthpressure.pdf}).
As discussed above \citep[and in further detail in][]{Pfeffer2018}, a higher birth pressure/density environment \footnote{Though density is the most relevant quantity when considering cluster mass loss, birth density and pressure are directly proportional (thus can be interchanged in analysis) due to the imposed equation of state at high gas densities in the EAGLE model.} leads to effective disruption of GCs and this is particularly prevalent in the low mass GCs, thus leading to a flattening of the mass function.
Interestingly, for $M_\ast > 10^{10.5} \Msun$ the power-law indices become smaller (steeper mass functions) with increasing galaxy mass, i.e. low-mass cluster mass loss is less effective in the higher mass galaxies, at odds with the observed increasing GC turnover mass with galaxy mass \citep{Jordan2007, Villegas2010}. This is found for all minimum mass assumptions in Fig.~\ref{fig:Mcstar_galmass_minmass.pdf}.
This may be caused by the increasing importance of galaxy mergers with increasing galaxy mass (as discussed in the previous section) resulting in a shorter time for cluster mass loss (Fig.~\ref{fig:galmass_TimeInParent.pdf}) before migration to a less disruptive environment (i.e. regions with weaker tidal fields).

Fig. \ref{fig:alpha_galmass_massloss.pdf} shows that the GC sample with no mass loss have a relatively consistent power law index, independently of galaxy mass. E-MOSAICS adopts $\alpha = -2$ for each star particle that forms a GC population. We would therefore expect that the subsample of GCs with no mass loss would have $\alpha = -2$, however Fig. \ref{fig:alpha_galmass_massloss.pdf} shows a steeper slope for this GC subsample. This is the effect of stacking many star particles (cluster subsamples), each with their own value for $\Mc$; the environmentally-dependent $\Mc$ means that those particles with lower $\Mc$ will contribute relatively more low-mass GCs, steepening the slope of the mass function.\par

Fig. \ref{fig:alpha_galmass_massloss.pdf} also shows that dynamical friction does not play a part in shaping the slope of the $z=0$ mass function, as expected. Dynamical friction time scales grow rapidly towards lower mass GCs  as $\tau_{\mathrm{DF}} \propto M_{\mathrm{GC}}^{-1}$ and therefore dynamical friction takes much longer to remove GCs with masses $M<\Mc$. By contrast, tidal shocks and two-body relaxation have a less obvious scaling with galaxy mass, because they depend on many galaxy properties. Therefore, although dynamical processes do shape the slope of the GC mass function, dynamical friction is not one of them.\par

\section{Conclusions \label{6}}
In this paper we investigate the origins of the shape of the high-mass end of the GC mass function as a function of galaxy stellar mass in a galaxy cluster. To carry out this analysis we have used the most massive galaxy group in the E-MOSAICS 34.4$^3$ cMpc$^3$ periodic volume. This choice was made to facilitate comparison to the observational results of \citet{Jordan2007} who fit evolved Schechter functions to GCs binned according to host galaxy mass in the Virgo Cluster. \par

Firstly we examine whether fitting Schechter functions is preferable over fitting power-law functions to the high-mass end of GC mass distributions. This is decided via a BIC test which penalises a maximum likelihood estimation based on the number of free parameters. We find that for the fiducial physics model in the simulations a Schechter function is preferable and therefore we can confidently compare the truncation mass of the Schechter function in the simulations to those of the observations in \citet{Jordan2007}. Fig. \ref{fig:galmass_fiducial} shows excellent agreement between the $\Mc$ of the simulations and the observations within the uncertainties.\par

To further investigate the input physics in the simulations we fit Schechter functions to the GCs that survive to $z=0$ under three different sets of cluster formation physics that do not allow the CFE, $\Mc$ or both, to vary with environment (Fig. \ref{fig:Mcstar_galmass_fiducial_altphys.pdf}). The model that does not let either the CFE or the $\Mc$ vary with environment still prefers a Schechter fit to a power-law fit because of dynamical friction disrupting the most massive GCs. However, this model yields an increasing $\Mc$ with galaxy mass due to a size-of-sample effect, and produces the wrong slope. The model that only allows the CFE to vary with environment again prefers a Schechter fit due to dynamical friction, but $\Mc$ remains too high to match observations. Finally, the model that only allows the $\Mc$ to vary with environment often prefers a power-law fit. This is because the high mass end of the mass function is not well sampled. Therefore it is only the fiducial model with an environmentally-dependent CFE and $\Mc$ that matches both the absolute values and the shape of the $\Mc$ trend with galaxy stellar mass. This adds to the body of work already supporting the initial physics and subsequent evolution of star clusters in the E-MOSAICS simulations. We therefore conclude that we can use the E-MOSAICS simulations to investigate the origin and shape of the observed trend.\par

\citet{Jordan2007} consider the argument that the decrease of $\Mc$ with decreasing galaxy mass could be due to the stronger depletion of massive GCs in dwarf galaxies due to dynamical friction. They postulate this because the dynamical friction timescale is proportional to the galaxy's circular speed ($\tau_{DF} \propto V_{\mathrm{c}}$) implying that the relevance of dynamical friction can increase in lower mass galaxies. They then rule out this hypothesis concluding that dynamical friction can only account for a small fraction of the steepening (decreasing $\Mc$) of the mass function with time. In the E-MOSAICS simulations, dynamical friction is applied in post-processing and therefore we can easily disable its effects and we do so in Fig. \ref{fig:Mcstar_galmass_massloss.pdf} and \ref{fig:alpha_galmass_massloss.pdf}. In the high-mass end of the GC mass function (Fig. \ref{fig:Mcstar_galmass_massloss.pdf}) we do not find that dynamical friction becomes more important in the lower-mass galaxies. In fact dynamical friction has a very small effect in these galaxies and the effect of dynamical friction on the slope of the GC mass function is negligible across all galaxy masses (Fig. \ref{fig:alpha_galmass_massloss.pdf}). Therefore our findings support the \citet{Jordan2007} claim that dynamical friction does not play an important role at low galaxy masses. However, at a galaxy stellar mass $\approx 10^{10} \Msun$ dynamical friction and other mass loss mechanisms are important in setting the shape of the high-mass end of the GC mass function (Fig. \ref{fig:Mcstar_galmass_massloss.pdf}).\par

GCs are formed with increasing $\Mc$ with galaxy mass until a galaxy mass $\log(M_{*}/M_{\odot})  \approx 10$ where $\Mc$ reaches a plateau and remains constant up to the largest masses. This is because the birth pressure also plateaus at these galaxy masses. The birth pressure plateaus because massive galaxies grow primarily by late mergers, and their GCs form in lower mass progenitors which have correspondingly lower pressures than the present day descendant.\par

Although $\Mc$ follows this trend at birth it is how the GCs are then nurtured by their parent galaxy that sets the final $\Mc$ we can observe today. This depends on whether the galaxy is able to move its high-mass GCs out of their highly disruptive birth environments quickly enough for them to survive until the present day. Galaxies can redistribute their GCs when they undergo merger events. Up until a stellar mass of $10^{10} \Msun$ galaxies are built by mainly in-situ star formation and therefore do not undergo enough mergers to re-distribute their GCs, leading to the destruction of their massive GCs. This means that disruption of all kinds is most efficient at reducing $\Mc$ at a stellar mass of around $10^{10}\Msun$ where the pressures/densities are high enough to form and then subsequently destroy high-mass GCs and there is not enough merger activity to redistribute them. \par

In conclusion, we find that it is a combination of both nature and nurture that sets the $z=0$ $\Mc$ trend with galaxy mass. It is in the galaxy's nature to form more massive GCs if the galaxy itself is massive, but these GCs must be nurtured and redistributed via mergers if they are to survive until $z=0$.

\section*{Acknowledgements}
JP and NB gratefully acknowledge funding from the ERC under the European Union’s Horizon 2020 research and innovation programme via the ERC Consolidator Grant Multi-Pop (grant agreement number 646928). JP gratefully acknowledges financial support from the Australian Research Council's Discovery Projects funding scheme (DP200102574). JMDK gratefully acknowledges funding from the Deutsche Forschungsgemeinschaft (DFG, German Research Foundation) through an Emmy Noether Research Group (grant number KR4801/1-1). JMDK and STG gratefully acknowledge funding from the European Research Council (ERC) under the European Union's Horizon 2020 research and innovation programme via the ERC Starting Grant MUSTANG (grant agreement number 714907). RAC is a Royal Society University Research Fellow. MRC gratefully acknowledges the Canadian Institute for Theoretical Astrophysics (CITA) National Fellowship for partial support. This study made use of high-performance computing facilities at Liverpool John Moores University, partly funded by the Royal Society and LJMUs Faculty of Engineering and Technology.
This work used the DiRAC Data Centric system at Durham University, operated by the Institute for Computational Cosmology on behalf of the STFC DiRAC HPC Facility (www.dirac.ac.uk). This equipment was funded by BIS National E-infrastructure capital grant ST/K00042X/1, STFC capital grants ST/H008519/1 and ST/K00087X/1, STFC DiRAC Operations grant ST/K003267/1 and Durham University. DiRAC is part of the National E-Infrastructure.

\section*{Data Availability}
The data underlying this article will be shared on reasonable request to the corresponding author.




\bibliographystyle{mnras}
\bibliography{mnras_template} 

\begin{thebibliography}{}
\makeatletter
\relax
\def\mn@urlcharsother{\let\do\@makeother \do\$\do\&\do\#\do\^\do\_\do\%\do\~}
\def\mn@doi{\begingroup\mn@urlcharsother \@ifnextchar [ {\mn@doi@}
  {\mn@doi@[]}}
\def\mn@doi@[#1]#2{\def\@tempa{#1}\ifx\@tempa\@empty \href
  {http://dx.doi.org/#2} {doi:#2}\else \href {http://dx.doi.org/#2} {#1}\fi
  \endgroup}
\def\mn@eprint#1#2{\mn@eprint@#1:#2::\@nil}
\def\mn@eprint@arXiv#1{\href {http://arxiv.org/abs/#1} {{\tt arXiv:#1}}}
\def\mn@eprint@dblp#1{\href {http://dblp.uni-trier.de/rec/bibtex/#1.xml}
  {dblp:#1}}
\def\mn@eprint@#1:#2:#3:#4\@nil{\def\@tempa {#1}\def\@tempb {#2}\def\@tempc
  {#3}\ifx \@tempc \@empty \let \@tempc \@tempb \let \@tempb \@tempa \fi \ifx
  \@tempb \@empty \def\@tempb {arXiv}\fi \@ifundefined
  {mn@eprint@\@tempb}{\@tempb:\@tempc}{\expandafter \expandafter \csname
  mn@eprint@\@tempb\endcsname \expandafter{\@tempc}}}

\bibitem[\protect\citeauthoryear{{Adamo} \& {Bastian}}{{Adamo} \&
  {Bastian}}{2018}]{Adamo2018}
{Adamo} A.,  {Bastian} N.,  2018, {The Lifecycle of Clusters in Galaxies}.
Springer International Publishing, Cham, p.~91,
  \mn@doi{10.1007/978-3-319-22801-3_4}

\bibitem[\protect\citeauthoryear{{Adamo}, {Kruijssen}, {Bastian}, {Silva-Villa}
   \& {Ryon}}{{Adamo} et~al.}{2015}]{Adamo2015M83}
{Adamo} A.,  {Kruijssen} J.~M.~D.,  {Bastian} N.,  {Silva-Villa} E.,   {Ryon}
  J.,  2015, \mn@doi [\mnras] {10.1093/mnras/stv1203}, \href
  {https://ui.adsabs.harvard.edu/abs/2015MNRAS.452..246A} {452, 246}

\bibitem[\protect\citeauthoryear{{Adamo} et~al.,}{{Adamo}
  et~al.}{2017}]{Adamo2017}
{Adamo} A.,  et~al., 2017, \mn@doi [\apj] {10.3847/1538-4357/aa7132}, \href
  {https://ui.adsabs.harvard.edu/abs/2017ApJ...841..131A} {841, 131}

\bibitem[\protect\citeauthoryear{{Adamo} et~al.,}{{Adamo}
  et~al.}{2020}]{Adamo2020}
{Adamo} A.,  et~al., 2020, \mn@doi [\ssr] {10.1007/s11214-020-00690-x}, \href
  {https://ui.adsabs.harvard.edu/abs/2020SSRv..216...69A} {216, 69}

\bibitem[\protect\citeauthoryear{{Bastian}, {Konstantopoulos}, {Trancho},
  {Weisz}, {Larsen}, {Fouesneau}, {Kaschinski}  \& {Gieles}}{{Bastian}
  et~al.}{2012}]{Bastian2012}
{Bastian} N.,  {Konstantopoulos} I.~S.,  {Trancho} G.,  {Weisz} D.~R.,
  {Larsen} S.~S.,  {Fouesneau} M.,  {Kaschinski} C.~B.,   {Gieles} M.,  2012,
  \mn@doi [\aap] {10.1051/0004-6361/201219079}, \href
  {https://ui.adsabs.harvard.edu/abs/2012A&A...541A..25B} {541, A25}

\bibitem[\protect\citeauthoryear{{Bastian}, {Pfeffer}, {Kruijssen}, {Crain},
  {Trujillo-Gomez}  \& {Reina-Campos}}{{Bastian} et~al.}{2020}]{Bastian2020}
{Bastian} N.,  {Pfeffer} J.,  {Kruijssen} J.~M.~D.,  {Crain} R.~A.,
  {Trujillo-Gomez} S.,   {Reina-Campos} M.,  2020, \mn@doi [\mnras]
  {10.1093/mnras/staa2453}, \href
  {https://ui.adsabs.harvard.edu/abs/2020MNRAS.498.1050B} {498, 1050}

\bibitem[\protect\citeauthoryear{{Baumgardt}}{{Baumgardt}}{1998}]{Baumgardt1998}
{Baumgardt} H.,  1998, \aap, \href
  {https://ui.adsabs.harvard.edu/abs/1998A&A...330..480B} {330, 480}

\bibitem[\protect\citeauthoryear{{Baumgardt}, {Parmentier}, {Anders}  \&
  {Grebel}}{{Baumgardt} et~al.}{2013}]{Baumgardt2013}
{Baumgardt} H.,  {Parmentier} G.,  {Anders} P.,   {Grebel} E.~K.,  2013,
  \mn@doi [\mnras] {10.1093/mnras/sts667}, \href
  {https://ui.adsabs.harvard.edu/abs/2013MNRAS.430..676B} {430, 676}

\bibitem[\protect\citeauthoryear{{Bik}, {Lamers}, {Bastian}, {Panagia}  \&
  {Romaniello}}{{Bik} et~al.}{2003}]{Bik2003}
{Bik} A.,  {Lamers} H.~J.~G.~L.~M.,  {Bastian} N.,  {Panagia} N.,
  {Romaniello} M.,  2003, \mn@doi [\aap] {10.1051/0004-6361:20021384}, \href
  {https://ui.adsabs.harvard.edu/abs/2003A&A...397..473B} {397, 473}

\bibitem[\protect\citeauthoryear{{Booth} \& {Schaye}}{{Booth} \&
  {Schaye}}{2009}]{Booth2009}
{Booth} C.~M.,  {Schaye} J.,  2009, \mn@doi [\mnras]
  {10.1111/j.1365-2966.2009.15043.x}, \href
  {https://ui.adsabs.harvard.edu/abs/2009MNRAS.398...53B} {398, 53}

\bibitem[\protect\citeauthoryear{{Brodie} \& {Strader}}{{Brodie} \&
  {Strader}}{2006}]{Brodie2006}
{Brodie} J.~P.,  {Strader} J.,  2006, \mn@doi [\araa]
  {10.1146/annurev.astro.44.051905.092441}, \href
  {https://ui.adsabs.harvard.edu/abs/2006ARA&A..44..193B} {44, 193}

\bibitem[\protect\citeauthoryear{{Burkert} \& {Smith}}{{Burkert} \&
  {Smith}}{2000}]{Burkert2000}
{Burkert} A.,  {Smith} G.~H.,  2000, \mn@doi [\apjl] {10.1086/312944}, \href
  {https://ui.adsabs.harvard.edu/abs/2000ApJ...542L..95B} {542, L95}

\bibitem[\protect\citeauthoryear{{Chandar} et~al.,}{{Chandar}
  et~al.}{2010}]{Chandar2010}
{Chandar} R.,  et~al., 2010, \mn@doi [\apj] {10.1088/0004-637X/719/1/966},
  \href {https://ui.adsabs.harvard.edu/abs/2010ApJ...719..966C} {719, 966}

\bibitem[\protect\citeauthoryear{{Chandar}, {Whitmore}, {Calzetti}  \&
  {O'Connell}}{{Chandar} et~al.}{2014}]{Chandar2014}
{Chandar} R.,  {Whitmore} B.~C.,  {Calzetti} D.,   {O'Connell} R.,  2014,
  \mn@doi [\apj] {10.1088/0004-637X/787/1/17}, \href
  {https://ui.adsabs.harvard.edu/abs/2014ApJ...787...17C} {787, 17}

\bibitem[\protect\citeauthoryear{{Chandar}, {Whitmore}, {Dinino}, {Kennicutt},
  {Chien}, {Schinnerer}  \& {Meidt}}{{Chandar} et~al.}{2016}]{Chandar2016}
{Chandar} R.,  {Whitmore} B.~C.,  {Dinino} D.,  {Kennicutt} R.~C.,  {Chien}
  L.~H.,  {Schinnerer} E.,   {Meidt} S.,  2016, \mn@doi [\apj]
  {10.3847/0004-637X/824/2/71}, \href
  {https://ui.adsabs.harvard.edu/abs/2016ApJ...824...71C} {824, 71}

\bibitem[\protect\citeauthoryear{{Chevance} et~al.,}{{Chevance}
  et~al.}{2020}]{Chevance2020}
{Chevance} M.,  et~al., 2020, \mn@doi [\ssr] {10.1007/s11214-020-00674-x},
  \href {https://ui.adsabs.harvard.edu/abs/2020SSRv..216...50C} {216, 50}

\bibitem[\protect\citeauthoryear{{Clauwens}, {Schaye}, {Franx}  \&
  {Bower}}{{Clauwens} et~al.}{2018}]{Clauwens2018}
{Clauwens} B.,  {Schaye} J.,  {Franx} M.,   {Bower} R.~G.,  2018, \mn@doi
  [\mnras] {10.1093/mnras/sty1229}, \href
  {https://ui.adsabs.harvard.edu/abs/2018MNRAS.478.3994C} {478, 3994}

\bibitem[\protect\citeauthoryear{{C{\^o}t{\'e}} et~al.,}{{C{\^o}t{\'e}}
  et~al.}{2004}]{Cote2004}
{C{\^o}t{\'e}} P.,  et~al., 2004, \mn@doi [\apjs] {10.1086/421490}, \href
  {https://ui.adsabs.harvard.edu/abs/2004ApJS..153..223C} {153, 223}

\bibitem[\protect\citeauthoryear{{Crain} et~al.,}{{Crain}
  et~al.}{2015}]{Crain2015}
{Crain} R.~A.,  et~al., 2015, \mn@doi [\mnras] {10.1093/mnras/stv725}, \href
  {http://adsabs.harvard.edu/abs/2015MNRAS.450.1937C} {450, 1937}

\bibitem[\protect\citeauthoryear{{Crain} et~al.,}{{Crain}
  et~al.}{2017}]{Crain2017}
{Crain} R.~A.,  et~al., 2017, \mn@doi [\mnras] {10.1093/mnras/stw2586}, \href
  {https://ui.adsabs.harvard.edu/abs/2017MNRAS.464.4204C} {464, 4204}

\bibitem[\protect\citeauthoryear{{Dalla Vecchia} \& {Schaye}}{{Dalla Vecchia}
  \& {Schaye}}{2012}]{DallaVecchia2012}
{Dalla Vecchia} C.,  {Schaye} J.,  2012, \mn@doi [\mnras]
  {10.1111/j.1365-2966.2012.21704.x}, \href
  {https://ui.adsabs.harvard.edu/abs/2012MNRAS.426..140D} {426, 140}

\bibitem[\protect\citeauthoryear{{Davison}, {Norris}, {Pfeffer}, {Davies}  \&
  {Crain}}{{Davison} et~al.}{2020}]{Davison2020}
{Davison} T.~A.,  {Norris} M.~A.,  {Pfeffer} J.~L.,  {Davies} J.~J.,   {Crain}
  R.~A.,  2020, \mn@doi [\mnras] {10.1093/mnras/staa1816}, \href
  {https://ui.adsabs.harvard.edu/abs/2020MNRAS.497...81D} {497, 81}

\bibitem[\protect\citeauthoryear{{Dowell}, {Buckalew}  \& {Tan}}{{Dowell}
  et~al.}{2008}]{Dowell2008}
{Dowell} J.~D.,  {Buckalew} B.~A.,   {Tan} J.~C.,  2008, \mn@doi [\aj]
  {10.1088/0004-6256/135/3/823}, \href
  {https://ui.adsabs.harvard.edu/abs/2008AJ....135..823D} {135, 823}

\bibitem[\protect\citeauthoryear{{Drinkwater}, {Gregg}  \&
  {Colless}}{{Drinkwater} et~al.}{2001}]{Drinkwater2001}
{Drinkwater} M.~J.,  {Gregg} M.~D.,   {Colless} M.,  2001, \mn@doi [\apjl]
  {10.1086/319113}, \href
  {https://ui.adsabs.harvard.edu/abs/2001ApJ...548L.139D} {548, L139}

\bibitem[\protect\citeauthoryear{{Durrell}, {Harris}, {Geisler}  \&
  {Pudritz}}{{Durrell} et~al.}{1996}]{Durrell1996}
{Durrell} P.~R.,  {Harris} W.~E.,  {Geisler} D.,   {Pudritz} R.~E.,  1996,
  \mn@doi [\aj] {10.1086/118071}, \href
  {https://ui.adsabs.harvard.edu/abs/1996AJ....112..972D} {112, 972}

\bibitem[\protect\citeauthoryear{{Elmegreen}}{{Elmegreen}}{2010}]{Elmegreen2010b}
{Elmegreen} B.~G.,  2010, \mn@doi [\apjl] {10.1088/2041-8205/712/2/L184}, \href
  {https://ui.adsabs.harvard.edu/abs/2010ApJ...712L.184E} {712, L184}

\bibitem[\protect\citeauthoryear{{Elmegreen}}{{Elmegreen}}{2011}]{Elmegreen2011}
{Elmegreen} B.~G.,  2011, in {Charbonnel} C.,  {Montmerle} T.,  eds,  EAS
  Publications Series Vol. 51, EAS Publications Series. pp 31--44 (\mn@eprint
  {arXiv} {1101.3111}), \mn@doi{10.1051/eas/1151003}

\bibitem[\protect\citeauthoryear{{Fall} \& {Zhang}}{{Fall} \&
  {Zhang}}{2001}]{Fall2001}
{Fall} S.~M.,  {Zhang} Q.,  2001, \mn@doi [\apj] {10.1086/323358}, \href
  {https://ui.adsabs.harvard.edu/abs/2001ApJ...561..751F} {561, 751}

\bibitem[\protect\citeauthoryear{{Fonnesbeck}, {Patil}, {Huard}  \&
  {Salvatier}}{{Fonnesbeck} et~al.}{2015}]{Fonnesbeck2015}
{Fonnesbeck} C.,  {Patil} A.,  {Huard} D.,   {Salvatier} J.,  2015, {PyMC:
  Bayesian Stochastic Modelling in Python} (\mn@eprint {ascl} {1506.005})

\bibitem[\protect\citeauthoryear{{Forbes} et~al.,}{{Forbes}
  et~al.}{2018}]{Forbes2018}
{Forbes} D.~A.,  et~al., 2018, \mn@doi [Proceedings of the Royal Society of
  London Series A] {10.1098/rspa.2017.0616}, \href
  {https://ui.adsabs.harvard.edu/abs/2018RSPSA.47470616F} {474, 20170616}

\bibitem[\protect\citeauthoryear{{Furlong} et~al.,}{{Furlong}
  et~al.}{2015}]{Furlong2015}
{Furlong} M.,  et~al., 2015, \mn@doi [\mnras] {10.1093/mnras/stv852}, \href
  {https://ui.adsabs.harvard.edu/abs/2015MNRAS.450.4486F} {450, 4486}

\bibitem[\protect\citeauthoryear{{Furlong} et~al.,}{{Furlong}
  et~al.}{2017}]{Furlong2017}
{Furlong} M.,  et~al., 2017, \mn@doi [\mnras] {10.1093/mnras/stw2740}, \href
  {https://ui.adsabs.harvard.edu/abs/2017MNRAS.465..722F} {465, 722}

\bibitem[\protect\citeauthoryear{Gieles, Zwart, Baumgardt, Athanassoula,
  Lamers, Sipior  \& Leenaarts}{Gieles et~al.}{2006a}]{Gieles2006b}
Gieles M.,  Zwart S. F.~P.,  Baumgardt H.,  Athanassoula E.,  Lamers H. J. G.
  L.~M.,  Sipior M.,   Leenaarts J.,  2006a, \mn@doi [Monthly Notices of the
  Royal Astronomical Society] {10.1111/j.1365-2966.2006.10711.x}, 371,
  793–804

\bibitem[\protect\citeauthoryear{{Gieles}, {Larsen}, {Bastian}  \&
  {Stein}}{{Gieles} et~al.}{2006b}]{Gieles2006}
{Gieles} M.,  {Larsen} S.~S.,  {Bastian} N.,   {Stein} I.~T.,  2006b, \mn@doi
  [\aap] {10.1051/0004-6361:20053589}, \href
  {https://ui.adsabs.harvard.edu/abs/2006A&A...450..129G} {450, 129}

\bibitem[\protect\citeauthoryear{{Goudfrooij}}{{Goudfrooij}}{2004}]{Goudfrooij2004}
{Goudfrooij} P.,  2004, in {Lamers} H. J.~G.~L.~M.,  {Smith} L.~J.,   {Nota}
  A.,  eds,  Astronomical Society of the Pacific Conference Series Vol. 322,
  The Formation and Evolution of Massive Young Star Clusters. p.~469
  (\mn@eprint {arXiv} {astro-ph/0404189})

\bibitem[\protect\citeauthoryear{{Gunn} \& {Gott}}{{Gunn} \&
  {Gott}}{1972}]{Gunn1972}
{Gunn} J.~E.,  {Gott} J.~Richard I.,  1972, \mn@doi [\apj] {10.1086/151605},
  \href {https://ui.adsabs.harvard.edu/abs/1972ApJ...176....1G} {176, 1}

\bibitem[\protect\citeauthoryear{{Hanes}}{{Hanes}}{1977}]{Hanes1977}
{Hanes} D.~A.,  1977, \mn@doi [\mnras] {10.1093/mnras/180.3.309}, \href
  {https://ui.adsabs.harvard.edu/abs/1977MNRAS.180..309H} {180, 309}

\bibitem[\protect\citeauthoryear{{Harris}}{{Harris}}{2001}]{Harris2001}
{Harris} W.~E.,  2001, in {Labhardt} L.,  {Binggeli} B.,  eds, Saas-Fee
  Advanced Course 28: Star Clusters. p.~223

\bibitem[\protect\citeauthoryear{{Harris} \& {Pudritz}}{{Harris} \&
  {Pudritz}}{1994}]{Harris1994}
{Harris} W.~E.,  {Pudritz} R.~E.,  1994, \mn@doi [\apj] {10.1086/174310}, \href
  {https://ui.adsabs.harvard.edu/abs/1994ApJ...429..177H} {429, 177}

\bibitem[\protect\citeauthoryear{{Harris} \& {Racine}}{{Harris} \&
  {Racine}}{1979}]{Harris1979}
{Harris} W.~E.,  {Racine} R.,  1979, \mn@doi [\araa]
  {10.1146/annurev.aa.17.090179.001325}, \href
  {https://ui.adsabs.harvard.edu/abs/1979ARA&A..17..241H} {17, 241}

\bibitem[\protect\citeauthoryear{{Hughes}, {Pfeffer}, {Martig}, {Bastian},
  {Crain}, {Kruijssen}  \& {Reina-Campos}}{{Hughes} et~al.}{2019}]{Hughes2019}
{Hughes} M.~E.,  {Pfeffer} J.,  {Martig} M.,  {Bastian} N.,  {Crain} R.~A.,
  {Kruijssen} J.~M.~D.,   {Reina-Campos} M.,  2019, \mn@doi [\mnras]
  {10.1093/mnras/sty2889}, \href
  {https://ui.adsabs.harvard.edu/abs/2019MNRAS.482.2795H} {482, 2795}

\bibitem[\protect\citeauthoryear{{Hughes}, {Pfeffer}, {Martig}, {Reina-Campos},
  {Bastian}, {Crain}  \& {Kruijssen}}{{Hughes} et~al.}{2020}]{Hughes2020}
{Hughes} M.~E.,  {Pfeffer} J.~L.,  {Martig} M.,  {Reina-Campos} M.,  {Bastian}
  N.,  {Crain} R.~A.,   {Kruijssen} J.~M.~D.,  2020, \mn@doi [\mnras]
  {10.1093/mnras/stz3341}, \href
  {https://ui.adsabs.harvard.edu/abs/2020MNRAS.491.4012H} {491, 4012}

\bibitem[\protect\citeauthoryear{{Johnson} et~al.,}{{Johnson}
  et~al.}{2017}]{Johnson2017}
{Johnson} L.~C.,  et~al., 2017, \mn@doi [\apj] {10.3847/1538-4357/aa6a1f},
  \href {https://ui.adsabs.harvard.edu/abs/2017ApJ...839...78J} {839, 78}

\bibitem[\protect\citeauthoryear{{Jord{\'a}n} et~al.,}{{Jord{\'a}n}
  et~al.}{2007}]{Jordan2007}
{Jord{\'a}n} A.,  et~al., 2007, \mn@doi [\apjs] {10.1086/516840}, \href
  {https://ui.adsabs.harvard.edu/abs/2007ApJS..171..101J} {171, 101}

\bibitem[\protect\citeauthoryear{{Kashibadze}, {Karachentsev}  \&
  {Karachentseva}}{{Kashibadze} et~al.}{2020}]{Kashibadze2020}
{Kashibadze} O.~G.,  {Karachentsev} I.~D.,   {Karachentseva} V.~E.,  2020,
  \mn@doi [\aap] {10.1051/0004-6361/201936172}, \href
  {https://ui.adsabs.harvard.edu/abs/2020A&A...635A.135K} {635, A135}

\bibitem[\protect\citeauthoryear{{Keller}, {Kruijssen}, {Pfeffer},
  {Reina-Campos}, {Bastian}, {Trujillo-Gomez}, {Hughes}  \& {Crain}}{{Keller}
  et~al.}{2020}]{Keller2020}
{Keller} B.~W.,  {Kruijssen} J.~M.~D.,  {Pfeffer} J.,  {Reina-Campos} M.,
  {Bastian} N.,  {Trujillo-Gomez} S.,  {Hughes} M.~E.,   {Crain} R.~A.,  2020,
  \mn@doi [\mnras] {10.1093/mnras/staa1439}, \href
  {https://ui.adsabs.harvard.edu/abs/2020MNRAS.495.4248K} {495, 4248}

\bibitem[\protect\citeauthoryear{{Kissler-Patig}, {Richtler}  \&
  {Hilker}}{{Kissler-Patig} et~al.}{1996}]{Kissler-Patig1996}
{Kissler-Patig} M.,  {Richtler} T.,   {Hilker} M.,  1996, \aap, \href
  {https://ui.adsabs.harvard.edu/abs/1996A&A...308..704K} {308, 704}

\bibitem[\protect\citeauthoryear{{Kruijssen}}{{Kruijssen}}{2012}]{Kruijssen2012}
{Kruijssen} J.~M.~D.,  2012, \mn@doi [\mnras]
  {10.1111/j.1365-2966.2012.21923.x}, \href
  {https://ui.adsabs.harvard.edu/abs/2012MNRAS.426.3008K} {426, 3008}

\bibitem[\protect\citeauthoryear{{Kruijssen}}{{Kruijssen}}{2014}]{Kruijssen2014}
{Kruijssen} J.~M.~D.,  2014, \mn@doi [Classical and Quantum Gravity]
  {10.1088/0264-9381/31/24/244006}, \href
  {https://ui.adsabs.harvard.edu/abs/2014CQGra..31x4006K} {31, 244006}

\bibitem[\protect\citeauthoryear{{Kruijssen}, {Pelupessy}, {Lamers}, {Portegies
  Zwart}  \& {Icke}}{{Kruijssen} et~al.}{2011}]{Kruijssen2011}
{Kruijssen} J.~M.~D.,  {Pelupessy} F.~I.,  {Lamers} H.~J.~G.~L.~M.,  {Portegies
  Zwart} S.~F.,   {Icke} V.,  2011, \mn@doi [\mnras]
  {10.1111/j.1365-2966.2011.18467.x}, \href
  {http://adsabs.harvard.edu/abs/2011MNRAS.414.1339K} {414, 1339}

\bibitem[\protect\citeauthoryear{{Kruijssen}, {Maschberger}, {Moeckel},
  {Clarke}, {Bastian}  \& {Bonnell}}{{Kruijssen}
  et~al.}{2012a}]{Kruijssen2012CC}
{Kruijssen} J.~M.~D.,  {Maschberger} T.,  {Moeckel} N.,  {Clarke} C.~J.,
  {Bastian} N.,   {Bonnell} I.~A.,  2012a, \mn@doi [\mnras]
  {10.1111/j.1365-2966.2011.19748.x}, \href
  {https://ui.adsabs.harvard.edu/abs/2012MNRAS.419..841K} {419, 841}

\bibitem[\protect\citeauthoryear{{Kruijssen}, {Pelupessy}, {Lamers}, {Portegies
  Zwart}, {Bastian}  \& {Icke}}{{Kruijssen} et~al.}{2012b}]{Kruijssen2012b}
{Kruijssen} J.~M.~D.,  {Pelupessy} F.~I.,  {Lamers} H. J.~G.~L.~M.,  {Portegies
  Zwart} S.~F.,  {Bastian} N.,   {Icke} V.,  2012b, \mn@doi [\mnras]
  {10.1111/j.1365-2966.2012.20322.x}, \href
  {https://ui.adsabs.harvard.edu/abs/2012MNRAS.421.1927K} {421, 1927}

\bibitem[\protect\citeauthoryear{{Kruijssen}, {Pfeffer}, {Crain}  \&
  {Bastian}}{{Kruijssen} et~al.}{2019a}]{Kruijssen2019a}
{Kruijssen} J.~M.~D.,  {Pfeffer} J.~L.,  {Crain} R.~A.,   {Bastian} N.,  2019a,
  \mn@doi [\mnras] {10.1093/mnras/stz968}, \href
  {https://ui.adsabs.harvard.edu/abs/2019MNRAS.486.3134K} {486, 3134}

\bibitem[\protect\citeauthoryear{{Kruijssen}, {Pfeffer}, {Reina-Campos},
  {Crain}  \& {Bastian}}{{Kruijssen} et~al.}{2019b}]{Kruijssen2019b}
{Kruijssen} J.~M.~D.,  {Pfeffer} J.~L.,  {Reina-Campos} M.,  {Crain} R.~A.,
  {Bastian} N.,  2019b, \mn@doi [\mnras] {10.1093/mnras/sty1609}, \href
  {https://ui.adsabs.harvard.edu/abs/2019MNRAS.486.3180K} {486, 3180}

\bibitem[\protect\citeauthoryear{{Kruijssen} et~al.,}{{Kruijssen}
  et~al.}{2020}]{Kruijssen2020}
{Kruijssen} J.~M.~D.,  et~al., 2020, \mn@doi [\mnras] {10.1093/mnras/staa2452},
  \href {https://ui.adsabs.harvard.edu/abs/2020MNRAS.498.2472K} {498, 2472}

\bibitem[\protect\citeauthoryear{{Krumholz}, {McKee}  \&
  {Bland-Hawthorn}}{{Krumholz} et~al.}{2019}]{Krumholz2019}
{Krumholz} M.~R.,  {McKee} C.~F.,   {Bland-Hawthorn} J.,  2019, \mn@doi [\araa]
  {10.1146/annurev-astro-091918-104430}, \href
  {https://ui.adsabs.harvard.edu/abs/2019ARA&A..57..227K} {57, 227}

\bibitem[\protect\citeauthoryear{{Lacey} \& {Cole}}{{Lacey} \&
  {Cole}}{1993}]{LaceyCole93}
{Lacey} C.,  {Cole} S.,  1993, \mn@doi [\mnras] {10.1093/mnras/262.3.627},
  \href {https://ui.adsabs.harvard.edu/abs/1993MNRAS.262..627L} {262, 627}

\bibitem[\protect\citeauthoryear{{Lagos} et~al.,}{{Lagos}
  et~al.}{2015}]{Lagos2015}
{Lagos} C. d.~P.,  et~al., 2015, \mn@doi [\mnras] {10.1093/mnras/stv1488},
  \href {https://ui.adsabs.harvard.edu/abs/2015MNRAS.452.3815L} {452, 3815}

\bibitem[\protect\citeauthoryear{{Larsen}}{{Larsen}}{2002}]{Larsen2002}
{Larsen} S.~S.,  2002, \mn@doi [\aj] {10.1086/342381}, \href
  {https://ui.adsabs.harvard.edu/abs/2002AJ....124.1393L} {124, 1393}

\bibitem[\protect\citeauthoryear{{Larsen}}{{Larsen}}{2009}]{Larsen2009}
{Larsen} S.~S.,  2009, \mn@doi [\aap] {10.1051/0004-6361:200811212}, \href
  {https://ui.adsabs.harvard.edu/abs/2009A&A...494..539L} {494, 539}

\bibitem[\protect\citeauthoryear{{Lee} \& {Yi}}{{Lee} \& {Yi}}{2013}]{Lee2013}
{Lee} J.,  {Yi} S.~K.,  2013, \mn@doi [\apj] {10.1088/0004-637X/766/1/38},
  \href {https://ui.adsabs.harvard.edu/abs/2013ApJ...766...38L} {766, 38}

\bibitem[\protect\citeauthoryear{{Liu}, {Peng}, {Jord{\'a}n}, {Blakeslee},
  {C{\^o}t{\'e}}, {Ferrarese}  \& {Puzia}}{{Liu} et~al.}{2019}]{Liu2019}
{Liu} Y.,  {Peng} E.~W.,  {Jord{\'a}n} A.,  {Blakeslee} J.~P.,  {C{\^o}t{\'e}}
  P.,  {Ferrarese} L.,   {Puzia} T.~H.,  2019, \mn@doi [\apj]
  {10.3847/1538-4357/ab12d9}, \href
  {https://ui.adsabs.harvard.edu/abs/2019ApJ...875..156L} {875, 156}

\bibitem[\protect\citeauthoryear{{Mackey} et~al.,}{{Mackey}
  et~al.}{2019}]{Mackey2019}
{Mackey} A.~D.,  et~al., 2019, \mn@doi [\mnras] {10.1093/mnras/stz072}, \href
  {https://ui.adsabs.harvard.edu/abs/2019MNRAS.484.1756M} {484, 1756}

\bibitem[\protect\citeauthoryear{{McCrady} \& {Graham}}{{McCrady} \&
  {Graham}}{2007}]{McCrady2007}
{McCrady} N.,  {Graham} J.~R.,  2007, \mn@doi [\apj] {10.1086/518357}, \href
  {https://ui.adsabs.harvard.edu/abs/2007ApJ...663..844M} {663, 844}

\bibitem[\protect\citeauthoryear{{McLaughlin}}{{McLaughlin}}{1999}]{McLaughlin1999}
{McLaughlin} D.~E.,  1999, \mn@doi [\apjl] {10.1086/311860}, \href
  {https://ui.adsabs.harvard.edu/abs/1999ApJ...512L...9M} {512, L9}

\bibitem[\protect\citeauthoryear{{Messa} et~al.,}{{Messa}
  et~al.}{2018a}]{Messa2018a}
{Messa} M.,  et~al., 2018a, \mn@doi [\mnras] {10.1093/mnras/stx2403}, \href
  {https://ui.adsabs.harvard.edu/abs/2018MNRAS.473..996M} {473, 996}

\bibitem[\protect\citeauthoryear{{Messa} et~al.,}{{Messa}
  et~al.}{2018b}]{Messa2018b}
{Messa} M.,  et~al., 2018b, \mn@doi [\mnras] {10.1093/mnras/sty577}, \href
  {https://ui.adsabs.harvard.edu/abs/2018MNRAS.477.1683M} {477, 1683}

\bibitem[\protect\citeauthoryear{{Miller}, {Whitmore}, {Schweizer}  \&
  {Fall}}{{Miller} et~al.}{1997}]{Miller1997}
{Miller} B.~W.,  {Whitmore} B.~C.,  {Schweizer} F.,   {Fall} S.~M.,  1997,
  \mn@doi [\aj] {10.1086/118655}, \href
  {https://ui.adsabs.harvard.edu/abs/1997AJ....114.2381M} {114, 2381}

\bibitem[\protect\citeauthoryear{{Mok}, {Chandar}  \& {Fall}}{{Mok}
  et~al.}{2019}]{Mok2019}
{Mok} A.,  {Chandar} R.,   {Fall} S.~M.,  2019, \mn@doi [\apj]
  {10.3847/1538-4357/aaf6ea}, \href
  {https://ui.adsabs.harvard.edu/abs/2019ApJ...872...93M} {872, 93}

\bibitem[\protect\citeauthoryear{{Okazaki} \& {Tosa}}{{Okazaki} \&
  {Tosa}}{1995}]{Okazaki1995}
{Okazaki} T.,  {Tosa} M.,  1995, \mn@doi [\mnras] {10.1093/mnras/274.1.48},
  \href {https://ui.adsabs.harvard.edu/abs/1995MNRAS.274...48O} {274, 48}

\bibitem[\protect\citeauthoryear{{Oser}, {Ostriker}, {Naab}, {Johansson}  \&
  {Burkert}}{{Oser} et~al.}{2010}]{Oser2010}
{Oser} L.,  {Ostriker} J.~P.,  {Naab} T.,  {Johansson} P.~H.,   {Burkert} A.,
  2010, \mn@doi [\apj] {10.1088/0004-637X/725/2/2312}, \href
  {https://ui.adsabs.harvard.edu/abs/2010ApJ...725.2312O} {725, 2312}

\bibitem[\protect\citeauthoryear{{Peng} et~al.,}{{Peng}
  et~al.}{2008}]{Peng2008}
{Peng} E.~W.,  et~al., 2008, \mn@doi [\apj] {10.1086/587951}, \href
  {https://ui.adsabs.harvard.edu/abs/2008ApJ...681..197P} {681, 197}

\bibitem[\protect\citeauthoryear{{Pfeffer}, {Kruijssen}, {Crain}  \&
  {Bastian}}{{Pfeffer} et~al.}{2018}]{Pfeffer2018}
{Pfeffer} J.,  {Kruijssen} J.~M.~D.,  {Crain} R.~A.,   {Bastian} N.,  2018,
  \mn@doi [\mnras] {10.1093/mnras/stx3124}, \href
  {http://adsabs.harvard.edu/abs/2018MNRAS.475.4309P} {475, 4309}

\bibitem[\protect\citeauthoryear{{Pfeffer}, {Bastian}, {Kruijssen},
  {Reina-Campos}, {Crain}  \& {Usher}}{{Pfeffer} et~al.}{2019}]{Pfeffer2019YC}
{Pfeffer} J.,  {Bastian} N.,  {Kruijssen} J.~M.~D.,  {Reina-Campos} M.,
  {Crain} R.~A.,   {Usher} C.,  2019, \mn@doi [\mnras] {10.1093/mnras/stz2721},
  \href {https://ui.adsabs.harvard.edu/abs/2019MNRAS.490.1714P} {490, 1714}

\bibitem[\protect\citeauthoryear{{Portegies Zwart}, {McMillan}  \&
  {Gieles}}{{Portegies Zwart} et~al.}{2010}]{PortegiesZwart2010}
{Portegies Zwart} S.~F.,  {McMillan} S. L.~W.,   {Gieles} M.,  2010, \mn@doi
  [\araa] {10.1146/annurev-astro-081309-130834}, \href
  {https://ui.adsabs.harvard.edu/abs/2010ARA&A..48..431P} {48, 431}

\bibitem[\protect\citeauthoryear{{Qu} et~al.,}{{Qu} et~al.}{2017}]{Qu2017}
{Qu} Y.,  et~al., 2017, \mn@doi [\mnras] {10.1093/mnras/stw2437}, \href
  {https://ui.adsabs.harvard.edu/abs/2017MNRAS.464.1659Q} {464, 1659}

\bibitem[\protect\citeauthoryear{{Racine}}{{Racine}}{1980}]{Racine1980}
{Racine} R.,  1980, in {Hesser} J.~E.,  ed., ~ Vol. 85, Star Clusters. pp
  369--380

\bibitem[\protect\citeauthoryear{{Reina-Campos} \& {Kruijssen}}{{Reina-Campos}
  \& {Kruijssen}}{2017}]{Reina-Campos2017}
{Reina-Campos} M.,  {Kruijssen} J.~M.~D.,  2017, \mn@doi [\mnras]
  {10.1093/mnras/stx790}, \href
  {https://ui.adsabs.harvard.edu/abs/2017MNRAS.469.1282R} {469, 1282}

\bibitem[\protect\citeauthoryear{{Reina-Campos}, {Kruijssen}, {Pfeffer},
  {Bastian}  \& {Crain}}{{Reina-Campos} et~al.}{2018}]{ReinaCampos2018}
{Reina-Campos} M.,  {Kruijssen} J.~M.~D.,  {Pfeffer} J.,  {Bastian} N.,
  {Crain} R.~A.,  2018, \mn@doi [\mnras] {10.1093/mnras/sty2451}, \href
  {https://ui.adsabs.harvard.edu/abs/2018MNRAS.481.2851R} {481, 2851}

\bibitem[\protect\citeauthoryear{{Reina-Campos}, {Kruijssen}, {Pfeffer},
  {Bastian}  \& {Crain}}{{Reina-Campos} et~al.}{2019}]{ReinaCampos2019}
{Reina-Campos} M.,  {Kruijssen} J.~M.~D.,  {Pfeffer} J.~L.,  {Bastian} N.,
  {Crain} R.~A.,  2019, \mn@doi [\mnras] {10.1093/mnras/stz1236}, \href
  {https://ui.adsabs.harvard.edu/abs/2019MNRAS.486.5838R} {486, 5838}

\bibitem[\protect\citeauthoryear{{Reina-Campos}, {Hughes}, {Kruijssen},
  {Pfeffer}, {Bastian}, {Crain}, {Koch}  \& {Grebel}}{{Reina-Campos}
  et~al.}{2020}]{ReinaCampos2020}
{Reina-Campos} M.,  {Hughes} M.~E.,  {Kruijssen} J.~M.~D.,  {Pfeffer} J.~L.,
  {Bastian} N.,  {Crain} R.~A.,  {Koch} A.,   {Grebel} E.~K.,  2020, \mn@doi
  [\mnras] {10.1093/mnras/staa483}, \href
  {https://ui.adsabs.harvard.edu/abs/2020MNRAS.493.3422R} {493, 3422}

\bibitem[\protect\citeauthoryear{{Rosas-Guevara} et~al.,}{{Rosas-Guevara}
  et~al.}{2015}]{Rosas-Guevara2015}
{Rosas-Guevara} Y.~M.,  et~al., 2015, \mn@doi [\mnras] {10.1093/mnras/stv2056},
  \href {https://ui.adsabs.harvard.edu/abs/2015MNRAS.454.1038R} {454, 1038}

\bibitem[\protect\citeauthoryear{{Schaye} \& {Dalla Vecchia}}{{Schaye} \&
  {Dalla Vecchia}}{2008}]{Schaye2008}
{Schaye} J.,  {Dalla Vecchia} C.,  2008, \mn@doi [\mnras]
  {10.1111/j.1365-2966.2007.12639.x}, \href
  {https://ui.adsabs.harvard.edu/abs/2008MNRAS.383.1210S} {383, 1210}

\bibitem[\protect\citeauthoryear{{Schaye} et~al.,}{{Schaye}
  et~al.}{2015}]{Schaye2015}
{Schaye} J.,  et~al., 2015, \mn@doi [\mnras] {10.1093/mnras/stu2058}, \href
  {http://adsabs.harvard.edu/abs/2015MNRAS.446..521S} {446, 521}

\bibitem[\protect\citeauthoryear{{Schechter}}{{Schechter}}{1976}]{Schechter1976}
{Schechter} P.,  1976, \mn@doi [\apj] {10.1086/154079}, \href
  {https://ui.adsabs.harvard.edu/abs/1976ApJ...203..297S} {203, 297}

\bibitem[\protect\citeauthoryear{{Schwarz}}{{Schwarz}}{1978}]{Schwarz1978}
{Schwarz} G.,  1978, Annals of Statistics, \href
  {https://ui.adsabs.harvard.edu/abs/1978AnSta...6..461S} {6, 461}

\bibitem[\protect\citeauthoryear{{Surdin}}{{Surdin}}{1979}]{Surdin1979}
{Surdin} V.~G.,  1979, \sovast, \href
  {https://ui.adsabs.harvard.edu/abs/1979SvA....23..648S} {23, 648}

\bibitem[\protect\citeauthoryear{{Toomre}}{{Toomre}}{1964}]{Toomre1964}
{Toomre} A.,  1964, \mn@doi [\apj] {10.1086/147861}, \href
  {https://ui.adsabs.harvard.edu/abs/1964ApJ...139.1217T} {139, 1217}

\bibitem[\protect\citeauthoryear{{Trayford} et~al.,}{{Trayford}
  et~al.}{2015}]{Trayford2015}
{Trayford} J.~W.,  et~al., 2015, \mn@doi [\mnras] {10.1093/mnras/stv1461},
  \href {https://ui.adsabs.harvard.edu/abs/2015MNRAS.452.2879T} {452, 2879}

\bibitem[\protect\citeauthoryear{{Trujillo-Gomez}, {Kruijssen}, {Reina-Campos},
  {Pfeffer}, {Keller}, {Crain}, {Bastian}  \& {Hughes}}{{Trujillo-Gomez}
  et~al.}{2021}]{TrujilloGomez2021}
{Trujillo-Gomez} S.,  {Kruijssen} J.~M.~D.,  {Reina-Campos} M.,  {Pfeffer}
  J.~L.,  {Keller} B.~W.,  {Crain} R.~A.,  {Bastian} N.,   {Hughes} M.~E.,
  2021, \mn@doi [\mnras] {10.1093/mnras/stab341}, \href
  {https://ui.adsabs.harvard.edu/abs/2021MNRAS.503...31T} {503, 31}

\bibitem[\protect\citeauthoryear{{Usher}, {Pfeffer}, {Bastian}, {Kruijssen},
  {Crain}  \& {Reina-Campos}}{{Usher} et~al.}{2018}]{Usher2018}
{Usher} C.,  {Pfeffer} J.,  {Bastian} N.,  {Kruijssen} J.~M.~D.,  {Crain}
  R.~A.,   {Reina-Campos} M.,  2018, \mn@doi [\mnras] {10.1093/mnras/sty1895},
  \href {https://ui.adsabs.harvard.edu/abs/2018MNRAS.480.3279U} {480, 3279}

\bibitem[\protect\citeauthoryear{{Vesperini}}{{Vesperini}}{1998}]{Vesperini1998}
{Vesperini} E.,  1998, \mn@doi [\mnras] {10.1046/j.1365-8711.1998.01837.x},
  \href {https://ui.adsabs.harvard.edu/abs/1998MNRAS.299.1019V} {299, 1019}

\bibitem[\protect\citeauthoryear{{Vesperini}, {Zepf}, {Kundu}  \&
  {Ashman}}{{Vesperini} et~al.}{2003}]{Vesperini2003}
{Vesperini} E.,  {Zepf} S.~E.,  {Kundu} A.,   {Ashman} K.~M.,  2003, \mn@doi
  [\apj] {10.1086/376688}, \href
  {https://ui.adsabs.harvard.edu/abs/2003ApJ...593..760V} {593, 760}

\bibitem[\protect\citeauthoryear{{Villegas} et~al.,}{{Villegas}
  et~al.}{2010}]{Villegas2010}
{Villegas} D.,  et~al., 2010, \mn@doi [\apj] {10.1088/0004-637X/717/2/603},
  \href {https://ui.adsabs.harvard.edu/abs/2010ApJ...717..603V} {717, 603}

\bibitem[\protect\citeauthoryear{{Whitmore} \& {Schweizer}}{{Whitmore} \&
  {Schweizer}}{1995}]{Whitmore1995}
{Whitmore} B.~C.,  {Schweizer} F.,  1995, \mn@doi [\aj] {10.1086/117334}, \href
  {https://ui.adsabs.harvard.edu/abs/1995AJ....109..960W} {109, 960}

\bibitem[\protect\citeauthoryear{{Whitmore}, {Zhang}, {Leitherer}, {Fall},
  {Schweizer}  \& {Miller}}{{Whitmore} et~al.}{1999}]{Whitmore1999}
{Whitmore} B.~C.,  {Zhang} Q.,  {Leitherer} C.,  {Fall} S.~M.,  {Schweizer} F.,
    {Miller} B.~W.,  1999, \mn@doi [\aj] {10.1086/301041}, \href
  {https://ui.adsabs.harvard.edu/abs/1999AJ....118.1551W} {118, 1551}

\bibitem[\protect\citeauthoryear{{Whitmore}, {Chandar}, {Bowers}, {Larsen},
  {Lindsay}, {Ansari}  \& {Evans}}{{Whitmore} et~al.}{2014}]{Whitmore2014}
{Whitmore} B.~C.,  {Chandar} R.,  {Bowers} A.~S.,  {Larsen} S.,  {Lindsay} K.,
  {Ansari} A.,   {Evans} J.,  2014, \mn@doi [\aj] {10.1088/0004-6256/147/4/78},
  \href {https://ui.adsabs.harvard.edu/abs/2014AJ....147...78W} {147, 78}

\bibitem[\protect\citeauthoryear{{Wiersma}, {Schaye}, {Theuns}, {Dalla Vecchia}
   \& {Tornatore}}{{Wiersma} et~al.}{2009}]{Wiersma2009}
{Wiersma} R. P.~C.,  {Schaye} J.,  {Theuns} T.,  {Dalla Vecchia} C.,
  {Tornatore} L.,  2009, \mn@doi [\mnras] {10.1111/j.1365-2966.2009.15331.x},
  \href {https://ui.adsabs.harvard.edu/abs/2009MNRAS.399..574W} {399, 574}

\bibitem[\protect\citeauthoryear{{Zhang} \& {Fall}}{{Zhang} \&
  {Fall}}{1999}]{Zhang1999}
{Zhang} Q.,  {Fall} S.~M.,  1999, \mn@doi [\apjl] {10.1086/312412}, \href
  {https://ui.adsabs.harvard.edu/abs/1999ApJ...527L..81Z} {527, L81}

\bibitem[\protect\citeauthoryear{{de Grijs}, {Bastian}  \& {Lamers}}{{de Grijs}
  et~al.}{2003}]{deGrijs2003}
{de Grijs} R.,  {Bastian} N.,   {Lamers} H. J.~G.~L.~M.,  2003, \mn@doi [\apjl]
  {10.1086/367928}, \href
  {https://ui.adsabs.harvard.edu/abs/2003ApJ...583L..17D} {583, L17}

\bibitem[\protect\citeauthoryear{{van den Bergh}}{{van den
  Bergh}}{1985}]{van_den_Bergh1985}
{van den Bergh} S.,  1985, \mn@doi [\apj] {10.1086/163535}, \href
  {https://ui.adsabs.harvard.edu/abs/1985ApJ...297..361V} {297, 361}

\makeatother
\end{thebibliography}





\appendix
\section{The fitting results under varying assumptions}\label{A}
The Schechter fits described in this work use a varying minimum mass which depends on the maximum GC mass in a particular galaxy mass bin. \citet{Jordan2007} also have a GC completeness luminosity that scales approximately with galaxy mass. In lower mass galaxies, lower mass GCs are more readily observable, due to the lower surface brightness of the galaxy. We choose a minimum GC mass to be 2 dex below the third most massive GC (to account for stochasticity at the high-mass end) but we now show the effect of using a constant minimum mass on the $\Mc$. 

\begin{figure}
    \centering
	\includegraphics[width=\linewidth]{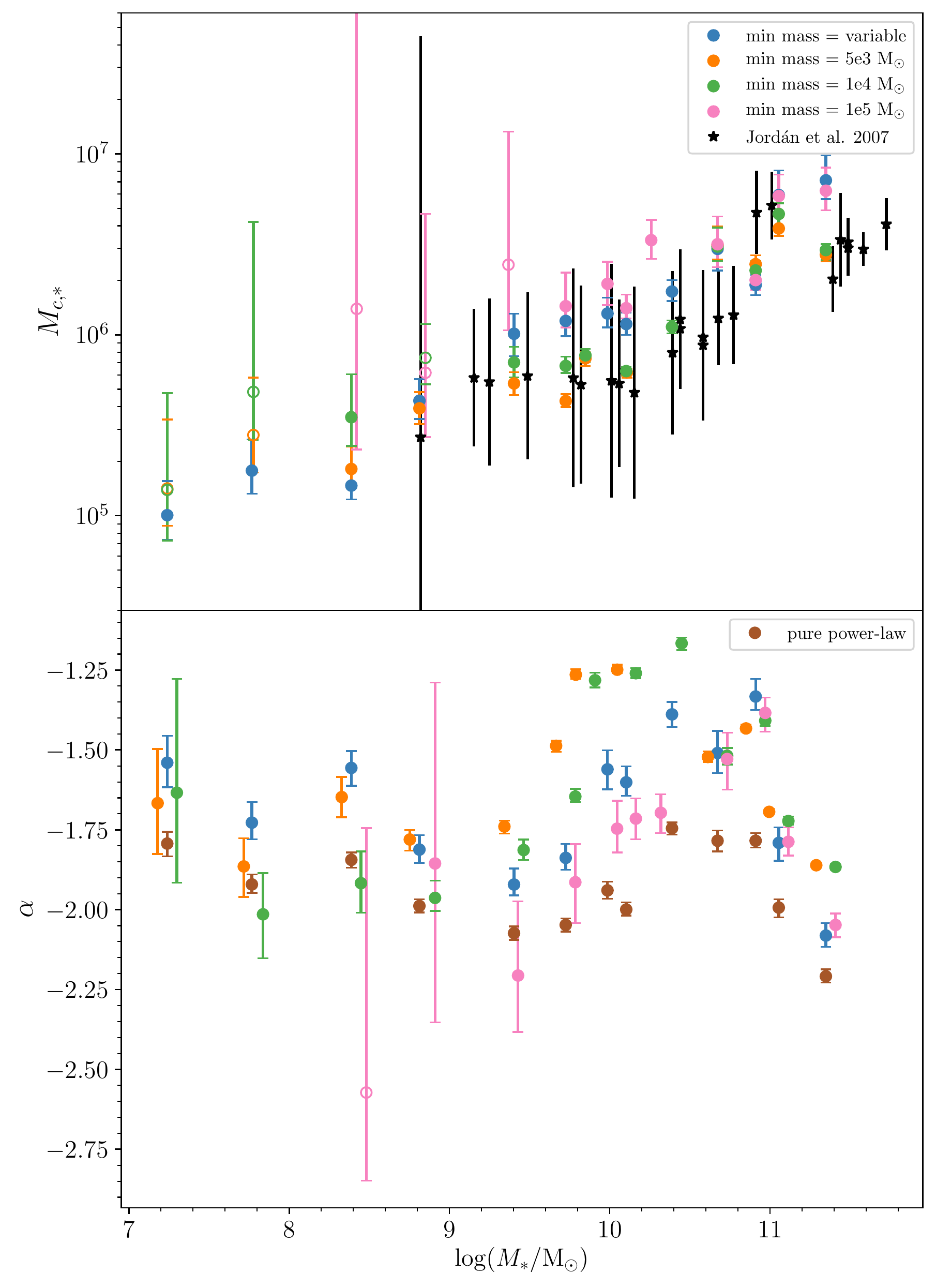}
    \caption{$\Mc$ and $\alpha$ as a function of galaxy mass for four different minimum mass cases. The variable case uses the minimum masses shown in table \ref{table:GCnumber}. The open symbols represent cases where the $\Mc$ from the Schechter fit is larger than the maximum GC mass in the galaxy bin. These are cases where a robust Schechter fit has not been found. In the lower panel we also show the slope obtained when fitting a pure power-law model, which is obtained by using a varying minimum mass.}
    \label{fig:Mcstar_galmass_minmass.pdf}
\end{figure}
In the upper panel of Fig.~\ref{fig:Mcstar_galmass_minmass.pdf} we compare the results of Schechter fits with varying minimum mass limits. The choice of minimum mass does somewhat affect the output of the Schechter fit (generally resulting in a slightly larger $\Mc$ with larger minimum mass), but the trend of $\Mc$ with $M_\ast$ remains similar. However, if the limit is too large then the window of masses we are fitting for some galaxies is too small for a robust fit. This is clear for the mass limit of $10^5 \Msun$ in the galaxies with $\log(M_*/\Msun) < 9.5$.  In many of these cases a Schechter fit could not be found and if one was outputted, the $\Mc$ was greater than the maximum GC mass (shown by open symbols).

In the lower panel of Fig.~\ref{fig:Mcstar_galmass_minmass.pdf} we present the value of the power-law index under these varying minimum mass assumptions.
For galaxy masses $M_\ast \lesssim 10^{9.5} \Msun$ the power-law indices obtained are relatively similar for all minimum mass assumptions, with the except for the $>10^5 \Msun$ limit which results in too few GCs for a robust fit at low galaxy masses.
For galaxies with $M_\ast > 10^{9.5} \Msun$ the power-law indices depend strongly on the lower mass limit (even though the recovered $\Mc$ is relatively insensitive). This indicate cluster mass loss is more important for these simulated galaxies, resulting in flatter mass functions at lower masses.
The indices obtained for a pure power-law fit are generally steeper than for Schechter models, which is expected given a single power-law function must also account for the steeper upper truncation of the mass functions.

\bsp	
\label{lastpage}
\end{document}